\documentclass[12pt,a4paper]{article}

\usepackage{amsmath,amssymb,bm,subfigure,float} 
\usepackage{dcolumn} 
\usepackage{bm} 
\usepackage{xcolor} 
\usepackage{psfrag} 
\usepackage[dvips]{graphicx} 
\begin{document}
 
\title{Interaction between Multi Components Vortices at Arbitrary Distances Using a Variational Method in the Ginzburg-Landau Theory}
 
\author{H.~Lookzadeh$^1$ and S.~Deldar$^2$ }


\abstract{
We study the interaction between the vortices in multi-component superconductors based on the Jacobs and Rebbi variation method using 
Ginzburg-Landau theory. With one condensation, we get attraction interaction between the vortices for type I and repulsion for type II 
superconductors. With two condensation states such as $MgB_2$ superconductors the behavior is quite different. There is attraction at large 
distances and repulsion when the vortices are close to each other. A stability point at distance $2.7/\lambda_1$ is obtained. In the case of three condensation 
states such as iron based superconductors, we see different behavior depending on penetration depth and correlation length. The formation 
energy of a vortex with three condensation states is larger than the one with one condensation state with comparable penetration and correlation length. We obtain two stability points for the superconductors with three condensation states. 
}
 
\maketitle 

\section{Introduction} 
The interaction between the elementary particles can be described by means of a field of force, just as the interaction between the charged particles which is described by the electromagnetic field. In quantum field theory, the electromagnetic field is accompanied by photon. Attraction and repulsion of electric charged particles can be described by exchanging particles called virtual photons \cite{Yukawa1935}. On the other hand topological defects are also important structures in physics since they can affect the properties of matter or even the phase structure of a system. These structures, such as vortices, monopoles, strings and instantons, can interact with each other like particles. They even have interaction with particles \cite{Nitta2014}. In this paper we study the interaction between the vortices in superconductor materials based on a variational numeric computation for arbitrary separation between vortices \cite{Manton,Kramer,Rebbi1979}. This method is useful in some 
 phenomenological models. Vortices are the solutions of the Ginzburg-Landau equations \cite{Ginzburg1950,Tinkham}. These equations give a topological structure with finite energy. The Ginzburg-Landau equations  are at least two nonlinear coupled equations, so there is no exact analytical solution for these equations. As the order of the nonlinearity is not small, it is not possible to use the usual perturbation methods to study the G-L Lagrangian behavior. Nevertheless, it is possible to study their behavior at asymptotic distances. Knowing the asymptotic behaviors of the functions, we can have an ansatz for these solutions for any arbitrary distances \cite{Rebbi1979}.

For the first time Abrikosov predicted the existence of a vortex structure in superconductors \cite{Abrikosov}. He suggested that the form of magnetic field penetration in a superconductor can be described by vortex equations. He studied the vortex properties by a Ginzburg-Landau theory. 
The G-L Lagrangian looks like an Abelian Higgs model where the Higgs field is like the order parameter and the gauge field is the electromagnetic field. The normal core of the vortex is introduced by the superconductor correlation length $ \xi $, and London penetration depth $\lambda$. 
There are two types of superconductors \cite{Annett,LinHu2011} depending on the Ginzburg-Landau parameter $\kappa=\lambda/\xi$. For $\kappa < 1/\sqrt{2}$, the magnetic penetration depth is smaller than the correlation length. This is the type I superconductor, for which the vortex structure is not stable. The interaction between vortices of this type is attraction. For $ \kappa > 1/\sqrt{2}$ , magnetic penetration depth is larger than the correlation length, this is type II superconductor. The magnetic field can penetrate in these materials. The vortex structure is stable. There is repulsion between these vortices and they form a triangular vortex lattice \cite{Annett,LinHu2011}.
 For $\kappa =1/ \sqrt{2}$ called the Bogomol'nyi point or type I/II border, there is no interaction between the vortices. 

Depending on the distance between the vortices, various methods can be chosen to study the interaction between them. 
Kramer \cite{Kramer} used asymptotic behavior of a vortex fields to obtain an analytical expression for the vortex-vortex interaction energy when they are far from each other. 
The fields can be explained by modified Bessel functions at the asymptotic regime;  
but what about the vortex interaction when they are close to each other? Jacobs and Rebbi used a variation method to obtain approximate trial functions describing the fields of two vortices at arbitrary distances \cite{Rebbi1979}.
 Variational parameters were obtained by minimizing the free energy. Their method can predict the  results of Kramer for large distances.
 It also predicts the same type of interaction for the small distances in type I and type II superconductors. There are other methods to study the interaction between the vortices \cite{Peeters2011}. 

A superconductor with more than one condensation state is called a multi-band superconductor.  
$MgB_2$ and iron pnictide superconductors are of this type. These materials have a higher phase transition temperature with respect to the usual superconductors of type I and II. 
They behave differently compared with the type I or type II superconductors with one condensation state.  
Also the possibility of the existence of more than three condensation states has been recently studied from the theoretical point of view \cite{Babaev and weston}. 
Interaction between these vortices is different from the usual superconductors \cite{multiband}.  
Babaev and Speight have studied theoretically \cite{Speight2005} what happens when the value of magnetic penetration depth is between
two condensation lengths; vortices may attract each other at large separations and repel each other at short distances.  
This kind of superconductor, called  type $1.5$ in the literature, is type I corresponding to one of its condensation and type II with respect to 
the other one. 

In this paper we use a G-L Lagrangian and the variational method of Jacobs and Rebbi to study the interaction between the vortices with three condensation
 states. The G-L theory is valid near the critical temperature. A different, coordinate system, the polar coordinate system, is used in our calculations. Since a single vortex has a circular symmetry or $SO(2)$ 
symmetry,  
choosing a polar coordinate system simplifies the calculations \cite{Manton}. However, when we have two vortices in a plane, we lose this symmetry and only a reflection symmetry 
with respect to the plane remains. The plane is located between the vortices.  First we use this method for a vortex with one condensation. Then we apply it for two and three 
condensations. The case with three condensations is different from the one with two condensations.  
The energy of formation of vortices of type $1.5$  is larger than the energy of type I and type II.  
Since the materials with multi-band condensation states are high temperature superconductors, the formation of these nonlinear 
structures with higher energy than usual superconductors may have some relations with the higher critical temperature in this kind of 
superconductors. Using this method of calculation one can suggest the values of the correlation lengths and penetration depths which increase 
the current known phase transition temperatures. 
 
\section{The Ginzburg-Landau Theory for Multiband Component Superconductor} 

The free energy of G-L theory can be given by  
\begin{equation}\label{eqsFF} 
E=\int{\mathcal{F} d \mathbf{r}}, 
\end{equation} 
where the functional $\mathcal{F}$ is 
\begin{equation}\label{eqsF1} 
\begin{array}{l} 
\mathcal{F}=\alpha \left |\Psi \right |{}^2+\frac{\beta }{2}\left |\Psi \right |{}^4+\frac{1}{2m}\left |\left(-i \hbar \nabla -\frac{2e} {c}\mathbf{A}\right)\Psi \right |{}^2 \\  
+\frac{1}{8 \pi}\left( \nabla \times \mathbf{A}\right)^2, 
\end{array} 
\end{equation}
the complex scalar field $\psi$ is the order parameter or the condensation state . $\mathbf{A}$ is a vector potential for magnetic field. 
$\alpha $ and $ \beta$ are the parameters that can be determined phenomenologically from the correlation length $\xi={\hbar}/{\sqrt{4m\alpha}}$ and the penetration depth 
$\lambda=\sqrt{mc^2\beta/8\pi|\alpha|c^2}$ of the superconducting matter \cite{Tinkham}. 
$\alpha$ is a temperature dependent parameter and is defined as $\alpha(T)=\alpha(0)(1-T/T_c)$ with $\alpha(0)<0$.  

One can generalize Eq. (\ref{eqsF1}) to a multi-band superconductor by increasing the number of condensation states. For example for 
two bands, the G-L theory can be introduced with two order parameters $|\psi|$ and for three bands with three states. In  the G-L theory one 
may consider other contributions up to $\psi^4$ terms. The contributions of  all types of possible interactions between fields in the G-L theory 
should be considered including $\psi_i \psi_j$ called interband coupling, $|\psi_i|^2|\psi_j|^2$ etc. The interband coupling terms, which 
do not exist in the usual superconductor, imply some new properties for the type $1.5$ superconductors. 
For the present work we consider only the interband coupling terms. The free energy functional for two condensation states is 
\begin{equation}\label{eqsF2} 
\begin{array}{l} 
\mathcal{F}=\\
\sum _{i=1,2}\left[\alpha _i\left|\Psi _i\right|^2+\frac{\beta _i}{2}\left|\Psi _i\right|^4+\frac{1}{2m_i}\left| \left(-i \hbar \nabla -\frac{2e}{c}\mathbf{A}\right)\Psi _i\right |^2 \right] \\
+\frac{1}{8\pi}(\nabla \times \mathbf{A} )^2-\gamma \left(\Psi _1^*\Psi _2+\Psi _2^*\Psi _1 \right),
\end{array} 
\end{equation}

where $\gamma(T)=\gamma(0)(1-T/T_c)$ (and $\gamma(0)=-0.4\alpha(0)$) is the condensations coupling.
The free energy functional for the case with three condensations is 
\begin{equation}\label{eqsF3} 
\begin{array}{l} 
\mathcal{F}=\\
\sum _{i=1,2,3}\left[\alpha _i\left |\Psi _i\right |{}^2+\frac{\beta _i}{2}\left |\Psi _i\right |{}^4+\frac{1}{2m_i}\left |\left(-i \hbar \nabla -\frac{2e}{c}\mathbf{A}\right)\Psi _i\right |{}^2\right]\\ 
+\frac{1}{8\pi }(\nabla \times \mathbf{A})^2
-\gamma_1 \left(\Psi _1^*\Psi _2+\Psi _2^*\Psi _1\right)-\gamma_2 \left(\Psi _2^*\Psi _3+\Psi _3^*\Psi _2\right)\\
-\gamma_3 \left(\Psi _1^*\Psi _3+\Psi _3^*\Psi _1\right). 
\end{array} 
\end{equation} 
For convenience we use the dimensionless quantities  
\begin{equation}\label{eqsFd} 
\begin{array}{l} 
x=\lambda _1x',\ \ \Psi _i=\Psi _{10}\Psi _i', \ \ \mathbf{A}=\lambda _1H_{1c}\sqrt{2}\mathbf{A}',\ \ \mathcal{F}=\frac{H_{1c}^2}{4\pi} \mathcal{F}',\\ 
\gamma =\gamma '\left |\alpha _1\right |,\ \ \mathbf{B}=H_{1c}\sqrt{2}\mathbf{B}',\ \ \mathbf{J}=\frac{2e\hbar \Psi _{10}^2}{m_1\xi _1}\mathbf{J}', 
\end{array} 
\end{equation} 
$\psi^2_{10}=|\alpha_1|/\beta_1$ is called the bulk value and $H_{1c}=\sqrt{4\pi\alpha_1\psi^2_{10}}$ is the thermodynamic critical field of the first condensate. $\mathbf{B}$ is the magnetic field and $\mathbf{J}$ is the super current. 
Omitting the prime for the dimensionless quantities, we have
\begin{equation}\label{eqsFd1} 
\begin{array}{l} 
\mathcal{F}=\\
\sum _{i=1,2,3}\left[\frac{\alpha _i}{\left |\alpha _1\right |}\left |\Psi _i\right |{}^2+\frac{\beta _i}{2\beta _1}\left |\Psi _i\right |{}^4+
\frac{m_1}{m_i}\left |\left(\frac{1}{i \kappa_1 }\nabla -\textbf{A}\right)\Psi _i\right |^2\right]\\
+(\nabla \times \textbf{A})^2 -\gamma_1 (\Psi _1^*\Psi _2+\Psi _2^*\Psi _1) \\
-\gamma_2 (\Psi _2^*\Psi _3+\Psi _3^*\Psi _2) -\gamma_3 (\Psi _1^*\Psi _3+\Psi _3^*\Psi _1). 
\end{array} 
\end{equation} 
The Euler-Lagrange equations can be obtained by
 
\begin{equation}\label{eqsEL} 
\begin{array}{c}
\dfrac{\partial\mathcal{F}}{\partial\Psi_\alpha}-\sum_i\dfrac{\partial}{\partial x_i}\dfrac{\partial\mathcal{F}}{\partial(\partial\Psi_\alpha/\partial x_i)}=0, \\
\dfrac{\partial\mathcal{F}}{\partial A_i}-\sum_i\dfrac{\partial}{\partial x_i}\dfrac{\partial\mathcal{F}}{\partial(\partial A_i/\partial x_i)}=0.
\end{array}
\end{equation} 
Solving these equations is not straightforward. One can use finite difference technique and a relaxation method 
suitable for nonlinear coupled differential equations to obtain the solutions which are used by Peeters \cite{Peeters2011}. Also it is possible to 
discretize the space and time with a method of lattice gauge theory to obtain the solutions \cite{LinHu2011}.
However, in this paper we use the variational method introduced by Jacobs and Rebbi to obtain trial functions for condensations and the vector potential. 
The advantage is that we can work analytically with these variational functions. However, it is a long analytical calculation. This method may be useful for studying the interaction between the vortices in the phenomenological models of particle physics which study the confinement problem \cite{thickvortex}. 

\section{Interaction between the vortices in type I and type II superconductors} 

We use the dimensionless free energy functional of (\ref{eqsF1}). Then the G-L equations from (\ref{eqsEL}) are obtained:
 \begin{equation}\label{eqsEM1} 
-\Psi +\left |\Psi \right |{}^2\Psi +\left(\frac{1}{i \kappa }\nabla -\textbf{A}\right){}^2\Psi =0, 
\end{equation} 
\begin{equation}\label{eqsEM2} 
\begin{array}{l} 
\nabla \times \nabla \times \textbf{A}=\frac{1}{2i \kappa _1}(\Psi ^*\nabla \Psi -\Psi \nabla \Psi ^*)-\left |\Psi \right |{}^2\textbf{A},
\end{array} 
\end{equation} 

where $\kappa_1=\lambda_{1}/\xi_{1}$.
 In Ref. \cite{Rebbi1979} a solution (ansatz) for $\psi$ and $ \vec{A}$ for the above equations is suggested:
 
\begin{equation}\label{eqsEM3} 
\Psi =f(r)e^{i n \theta } \text{  and } \textbf{A}=\frac{n a(r)}{\kappa _1 r}\textbf{e}_{\theta }. 
\end{equation} 
These are true for a straight vortex line type structure along the axis $z$. $r$ is the distance from the center of the vortex core. $\mathbf{e}_\theta$ is the unit 
vector, $\theta$ is the azimuthal direction, and $n$ represents the vorticity or the winding number. 
It is natural to discuss these circularly symmetric solutions in polar coordinates. Thus the fields are $\Psi(r,\theta)$,$A_{r}(r,\theta)$ and $A_{\theta}(r,\theta)$. We shall use the circular and 
reflection symmetries to obtain a reduced GL energy function, an integral just over the radial coordinate $r$. The variational equations are the reduced field equations. By the principle of 
symmetric criticality described in \cite{Manton}, solutions of these reduced equations give solutions of the full field equations in the plane.
Substituting (\ref{eqsEM3}) in (\ref{eqsEM1}) and  (\ref{eqsEM2})  
 \begin{equation}\label{eqsEMA1} 
-f(r)+f^3(r)-\frac{1}{\kappa_1 ^2}\left(\partial _r^2f+\frac{1}{r}\partial _rf\right)+\frac{n^2( a-1)^2}{\kappa_1 ^2r^2}f=0, 
\end{equation} 
\begin{equation}\label{eqsEMA2} 
\partial _r^2a-\frac{1}{r}\partial _ra+(f^2)(1-a)=0. 
\end{equation} 
The asymptotic forms of $f$ and $a$ for these explicit expressions do exist. We define the functions $F$ and $G$ such that 
\begin{equation}\label{F G}
f(r)=1+F(r)\;\;,\;\;\; a(r)=1+G(r)
\end{equation}
where $F$ and $G$ are small at large $r$. Thus substituting \ref{F G} in \ref{eqsEMA1} and \ref{eqsEMA2} and linearizing with respect to $F$ and $G$ one would get modified Bessel's 
equations of zeroth order for $F$ as a function of $k_{1}r$ and first order for $G/r$ as a function of $r$, respectively. Hence for $r \gg 1$
\begin{equation}
F \approx K_{0}(\sqrt{2} k_{1}r), \;\;\;\;\; G \approx r K_{1}(r),
\end{equation}
where $ K_{n}$ is the nth modified Bessel's function of the second kind (note that $K_{1}=-K^{'}_{0}$).
Solutions exist for any $N \neq 0 $ and can be found numerically. Near $r=0$, $f(r) \approx r^{N}$. 
From the above equations, the asymptotic values of $f$ and $a$ , $f_0$ and $a_0$, for $r\rightarrow\infty$ are obtained:
\begin{equation} 
-1+f_{0}^2=0\; \;, \; \; a_{0}=1. 
\end{equation} 
The radial variation of the wave functions and vector potential in the asymptotic region of $r\rightarrow\infty$ can be found and are given by
\begin{equation}\label{eqsTF10} 
f(r)=1+c_{f_{1}}\exp \left(-\frac{r}{\sqrt{2}\xi}\right), 
\end{equation} 
\begin{equation}\label{eqsTF20} 
a(r)=1+c_{a}\exp \left(-\frac{r}{\lambda}\right) .
\end{equation} 
where $c_{f_{1}}$and $c_{a}$ are the coefficients that can be found. Nielsen and Olesen \cite{Abrikosov} obtained similar solutions at the asymptotic region. These are also called Nielsen-Olesen solutions. 
Having the asymptotic behavior of the solutions at $r=0$ and $r \rightarrow \infty$ one would suggest an acceptable fitting functions that would recover these asymptotic, and can give acceptable intermediate behavior.
A polynomial times an exponential would be a good solution. The coefficients of the polynomial must be obtained numerically. Jacobs and Rebbi used a variational method to obtain these 
coefficients. To obtain $f(r)$, variational functions are suggested \cite{Rebbi1979} and the asymptotic behaviors of $f(r)$ and $a(r)$ fix the variational 
parameters: $f_l$ and $a_l$

\begin{equation}\label{eqsTF1} 
f(r)=1+\exp \left(-\frac{r}{\sqrt{2}\xi}\right)\sum _{l=0}^n\left(f_{l}\left.r^l\right/l!\right), 
\end{equation} 
\begin{equation}\label{eqsTF2} 
a(r)=1+\exp \left(-\frac{r}{\lambda}\right)\sum _{l=0}^n\left(a_l\left.r^l\right/l!\right) .
\end{equation} 
To have single-value and finite functions for $\psi$ and $\vec{A}$ in the limit of $ r \rightarrow 0 $, we use $f=0$ and $ a^2 \rightarrow 0 $. Note
 that in this method the G-L equations are not solved directly but by using their asymptotic behavior, we suggest some trial functions 
which minimize the free energy. The trial functions which minimize the free energy are solutions of the G-L equations, as well. For a vortex with 
vorticity two, asymptotic behavior gives to $ f_1=f_0/{\sqrt{2}\xi} $ and all other coefficients are variational parameters 
which are determined numerically.
 
The G-L free energy of Eq. (\ref{eqs12F1}) is a function of fourth order with respect to variational parameters, called $V_i$ in the following 
equation:  
\begin{equation}\label{eqsFV} 
\begin{array}{l} 
\mathcal{F}=\mathcal{F}_0+\sum _i\mathcal{F}_i^{(1)}V_i+\sum _{i\geq j}\mathcal{F}_{{ij}}^{(2)}V_iV_j+\\
\sum _{i\geq j\geq k}\mathcal{F}_{{ijk}}^{(3)}V_iV_jV_k+\sum _{i\geq j\geq k\geq l}\mathcal{F}_{{ijkl}}^{(4)}V_iV_jV_kV_l. 
\end{array} 
\end{equation}

We recall that in our problem the variational parameters are $f_i$ and $a_i$. The physical nature of the problem makes the surface 
$\mathcal{F}(f_i,a_i)$ concave and well behaved, so we use the Newton method of optimization \cite{Rebbi1979,LinHu2011} 
with iteration procedure
 
\begin{equation}\label{eqsITP} 
V_i^{(m+1)}=V_i^{(m)}-\sum _j\left[\mathbf{H}^{-1}\right]{}_{ij}D_j^{(m)}, 
\end{equation} 

$\mathbf{H}$ is the Hessian matrix and $D_i=\partial \mathcal{F}\left/\partial V _{i}\right.|_{V_i=V_{i}^{(m)}}$. 
The stationary solution of this equation corresponds to the (local) minimum of the free energy. In our computations, changing initial values
 of $V_i$ in the program does not change the obtained values of $V_i$, so the solutions  correspond to the absolute minimum of the free energy. 
We use this method to obtain the variational parameters for vorticity one and two.

We use this variational method to calculate the variational parameters of the condensation states and magnetic field for three types of 
superconductors: type I for which $ \xi =51nm$ and $ \lambda =25nm $ and $\kappa<1/\sqrt{2}$; type II for which $ \xi =19nm$ and 
$ \lambda =25nm$ and $\kappa>1/\sqrt{2}$; the Bogomoliny point where $ \xi =35nm$, $ \lambda =25nm$ and $ \kappa =1/\sqrt{2}$.
 For these types of superconductors we use the variational parameters up to the $8th$ polynomial terms. Using more terms and 
parameters does not change the free energy value up to the decimal point.    
\begin{figure*}[ht] 
\begin{minipage}[b]{0.4\linewidth} 
\centering 
\subfigure{(a)$ \kappa =0.49$}\\ 
{ 
\includegraphics[width=1\linewidth]{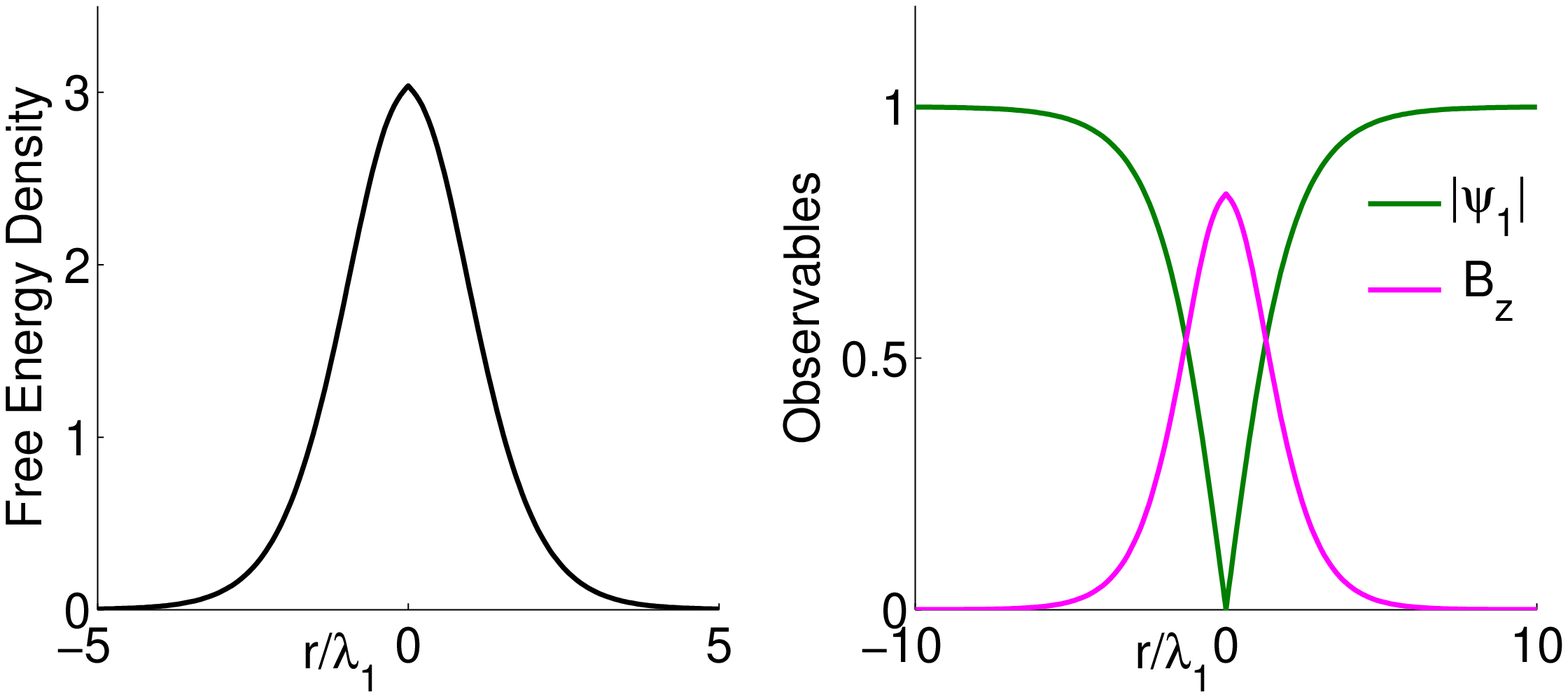} 
\label{Onecon1a} 
} 
\\ 
\subfigure{(b)$\kappa=0.71$}\\ 
{
\includegraphics[width=1\linewidth]{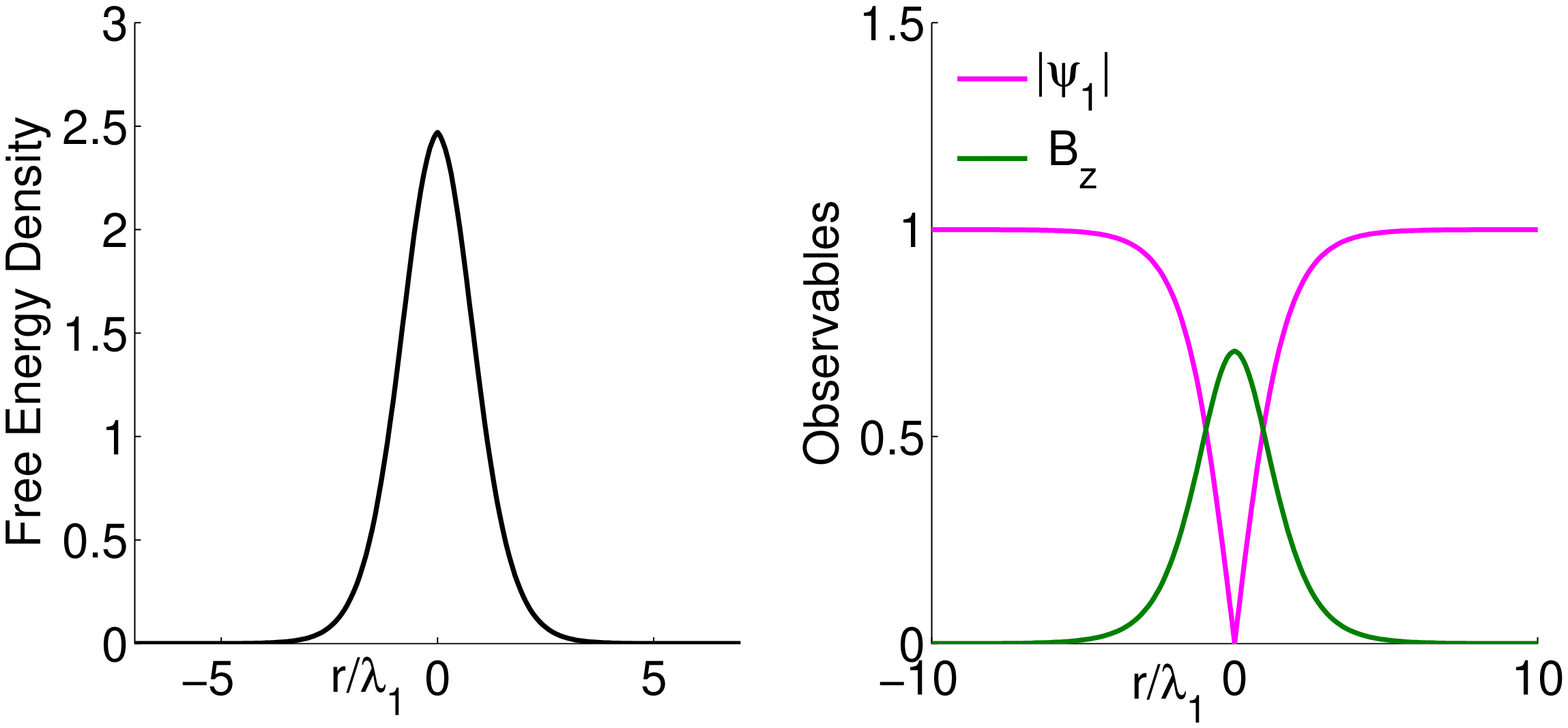} 
\label{Onecon1b} 
}
\\ 
\subfigure{(c)$\kappa=1.3$}\\
{ 
\includegraphics[width=1\linewidth]{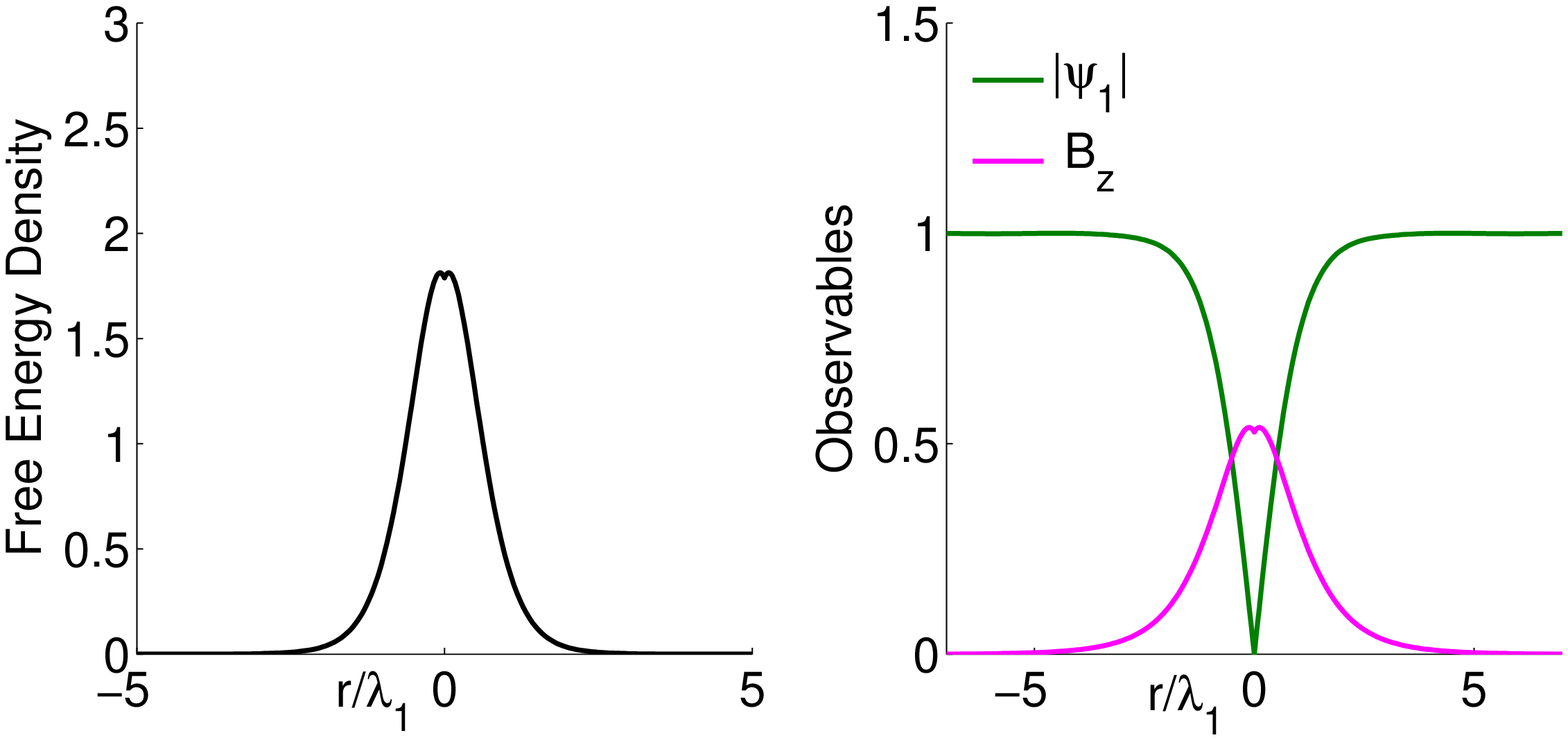} 
\label{Onecon1c} 
} 
\caption{ Free energy density and condensation state and magnetic field profiles cross sections in a plane for three types of G-L parameter with winding 
number $n=1$. $k=0.49$ is for the type I superconductor. $k=1.3$ is for the type II superconductor. The bulk behaviors of condensations and magnetic fields 
are the same for all three cases, as expected. Increasing the G-L parameter leads to a faster screening of the condensation and the magnetic 
field penetration depth decreases. The free energy value decreases by increasing the G-L parameter, as well. } 
\label{Onecon1} 
\end{minipage}
\hspace{1 cm} 
\begin{minipage}[b]{0.4\linewidth} 
\centering 
\subfigure{(2a)$\kappa=0.49$}
\\ 
{ 
\includegraphics[width=1\linewidth]{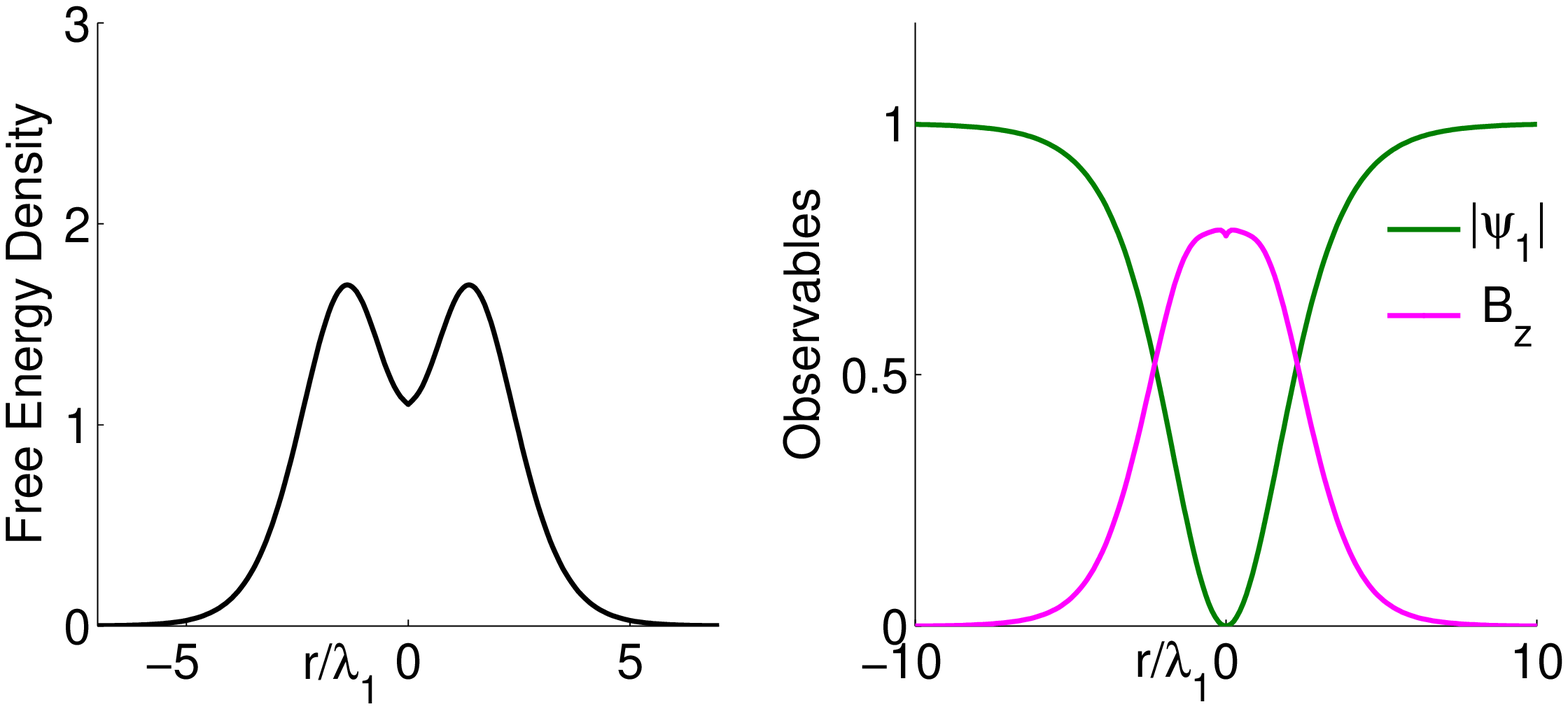} 
\label{Onecon2a} 
}
\\ 
\subfigure{(2b)$\kappa=0.71$}\\ 
{ 
\includegraphics[width=1\linewidth]{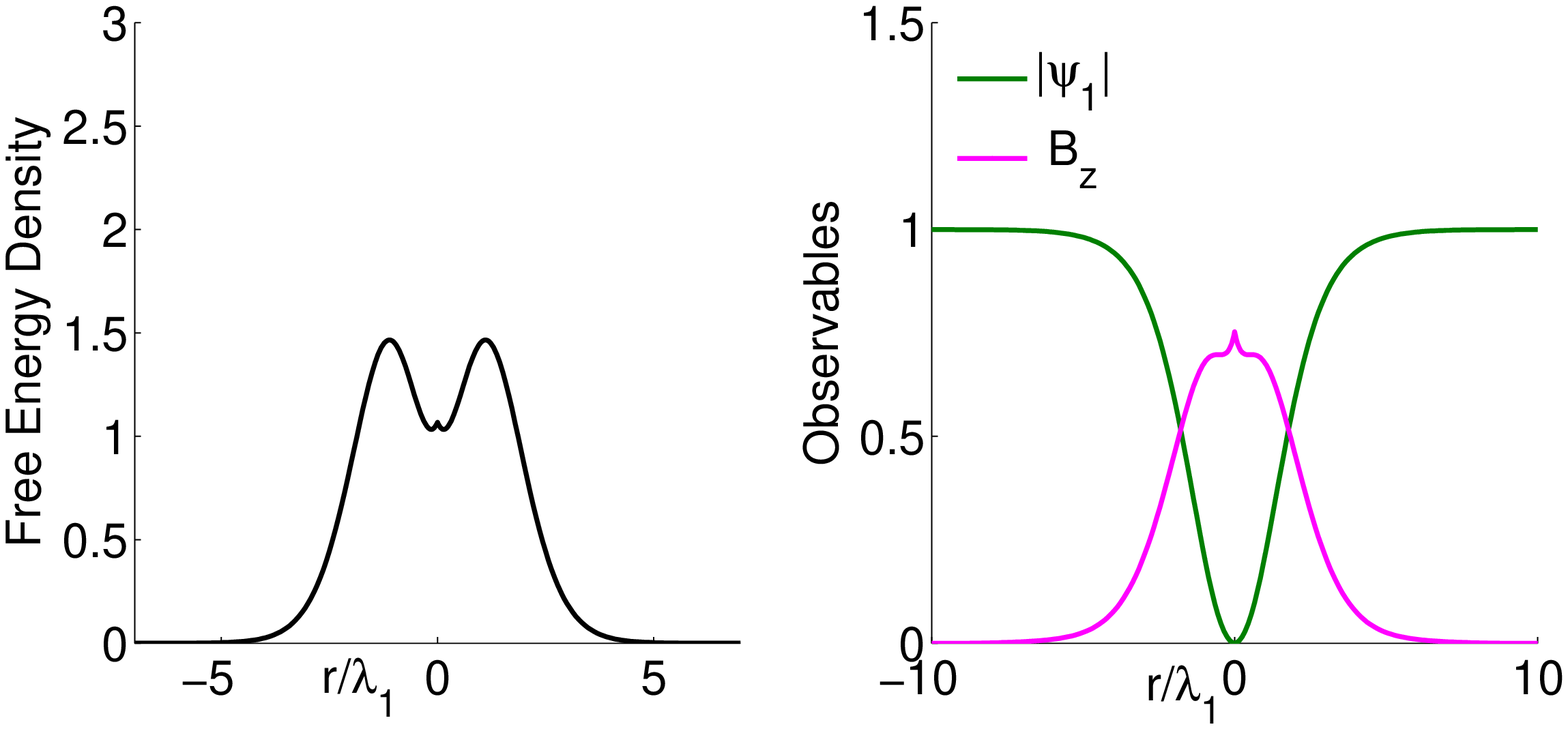} 
\label{Onecon2b} 
} 
\\ 
\subfigure{(2c)$\kappa=1.3$}\\ 
{
\includegraphics[width=1\linewidth]{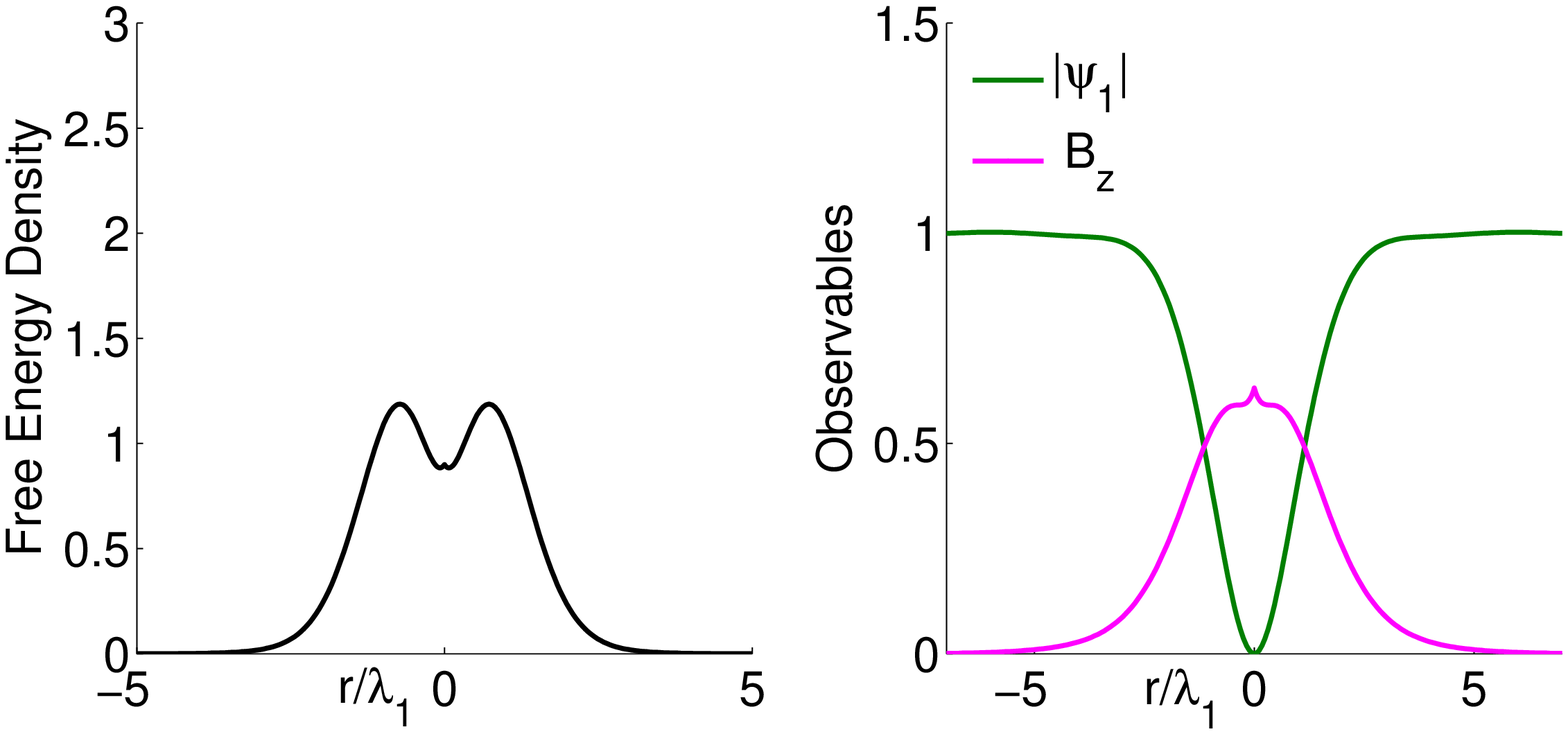} 
\label{Onecon2c} 
} 
\caption{Free energy density and condensation state and magnetic field profiles cross sections in a plane for the same G-L parameter as for Fig.(\ref{Onecon1}a-c)  for 
winding number $n=2$. The energy of a vortex formation decreases with increasing G-L parameter in the defined dimension of energy. 
Increasing the winding number to $n=2$, the maximum of the free energy density leaves the origin core of the vortex to other values. The results of 
our program show the same behavior for the free energy density as in \cite{Manton}, which shows that by increasing the winding number, the maximum of the free energy density is located at further distances from the origin place of the core of the vortex.} 
\label{Onecon2} 
\end{minipage} 
\end{figure*} 

\begin{figure}
\centering 
\includegraphics[width=0.5\linewidth]{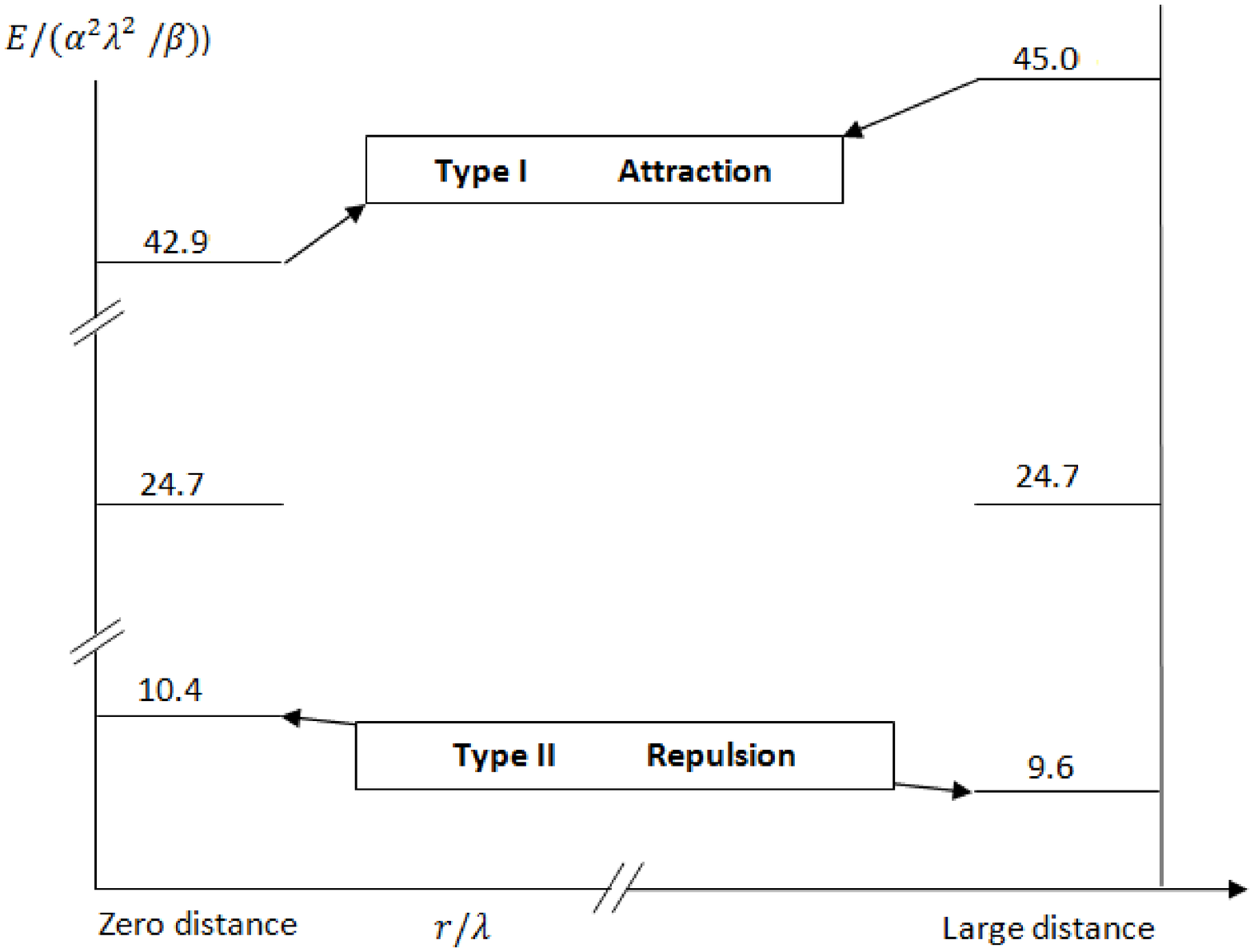} 
\caption{Energy versus distance between two vortices of type I and II when they are far from each other and when they merge and form one giant vortex with winding $n=2$, using the Jacobs and Rebbi variational method. For type I the energy of the giant vortex is smaller than two 
separate vortices, so the interaction is attraction. For type II the situation is reversed and the interaction is repulsion. For the 
case $\kappa=1/\sqrt{2}$ there is no interaction between the vortices.} 
\label{fig3} 
\end{figure}

Figure \ref{Onecon1} shows the condensation and magnetic field behaviors and also the free energy density for vorticity one for these three 
types. 
The free energies are $24.7,4.8$ and $12.3$, respectively. 
Figure (\ref{Onecon2}) shows the same functions for vorticity two for the same parameters of superconductor types of Fig. \ref{Onecon1}a-c.
The free energies for winding number $n=2$ are $42.9,10.4$ and $24.7$, respectively. The dimension of energy is $E/(\alpha^2\lambda^2/\beta)$. 
The free energy of a system consisting of two vortices located far from each other is equal to the sum of the free energy of two separate 
vortices with vorticity one. 
When they merge at zero distance, the energy is equal to the energy of a vortex with vorticity two. 
Therefore, if the energy of a vortex with vorticity two is larger than energy of two vortices with vorticity one, the interaction is repulsion and 
if the energy of a vortex with vorticity two is smaller than energy of two vortices with vorticity one, the interaction is attraction. 
Our results are $12.36828909$ and $24.76771608$ for $n=1$ and $n=2$, respectively, at the Bogomol'iny point.
For the Bogomol'iny point there is no interaction, As $24.76771608-2 \times 12.36828909=0.0311379$ the meaningful number of our calculation is up to the decimal point in this dimension of energy. Our results for a system of two vortices are shown in Fig. \ref{fig3}, a repulsion for type II and an attraction for type I are observed and for $\kappa=1/\sqrt{2}$ the vortices do not interact with each other. The results of \cite{Rebbi1979} also show a monotonic type interaction type, attraction and repulsion between the vortices for type I and type II superconductors, respectively, at any arbitrary distances.

\section{Interaction between the vortices in type $1.5$ superconductor} 

In this case the magnetic penetration depth lies between the two correlation lengths. Therefore, the interaction type of attraction or repulsion 
is not clear by obtaining the asymptotic value of free energy. This is called the type $1.5$ superconductor. Therefore, the variational method which has 
been used in the previous section must be applied considering details for this type of superconductor to study the interactions for all distances. 
The free energy is of (\ref{eqsF2}) type. For simplicity, it is assumed that the phase transition temperature is the same for both condensations. 
The G-L equations become 
\begin{equation}\label{eqs12v1} 
-\Psi _1+\left |\Psi _1\right |{}^2\Psi _1+\left(\frac{1}{i \kappa _1}\nabla -\textbf{A}\right){}^2\Psi _1-\gamma \Psi _2-=0, 
\end{equation} 
\begin{equation}\label{eqs12v2} 
-\frac{\alpha _2}{\alpha _1}\Psi _2+\frac{\beta _2}{\beta _1}\left |\Psi _2\right |{}^2\Psi _2+\frac{m_1}{m_2}\left(\frac{1}{i \kappa _1}\nabla -\textbf{A}\right){}^2\Psi _2-\gamma \Psi _1=0, 
\end{equation} 
\begin{equation}\label{eqs12v3} 
\begin{array}{l} 
\nabla \times \nabla \times \textbf{A}=\frac{1}{2i \kappa _1}(\Psi _1^*\nabla \Psi _1-\Psi _1\nabla \Psi _1^*)-\left |\Psi _1\right |{}^2\textbf{A} \\
+\frac{m_1}{m_2}\left(\frac{1}{2i \kappa _1}(\Psi _2^*\nabla \Psi _2-\Psi _2\nabla \Psi _2^*)-\left |\Psi _2\right |{}^2\textbf{A}\right). 
\end{array} 
\end{equation} 
Applying the London approximation to Eq. (\ref{eqs12v3}), one gets to an effective London penetration depth for the two-bands superconductor: 
\begin{equation}\label{eqs12v4} 
\lambda =1/\left(\sqrt{\left |\Psi _{10}\right |{}^2+\frac{m_1}{m_2}\left |\Psi _{20}\right |{}^2} \right) .
\end{equation} 
For $\gamma>0$, called positive coupling coefficient, the two condensates must have the same vorticity \cite{LinHu2011}. $MgB_2$ is an 
example of this kind of superconductors.  $\lambda$ is going to be used in the trial function of vector potential.
When the winding numbers for the two condensations are not equal, the flux of the vortex is fractionally quantized and the energy diverges 
logarithmically \cite{Babaev2007}. These are not topologically stable structures. Throughout this article we do not consider these fractional vortices or a non-topological one.
It is not possible to use the variational method for the situations when the phase or winding of all condensations are not equal.

Using the same vortex line ansatz of section III:
\begin{equation}\label{eqs12v5} 
\Psi _i=f_i(r)e^{i n \theta } \text{\ and\ } \textbf{A}=\frac{n a(r)}{\kappa _1 r}\textbf{e}_{\theta }. 
\end{equation} 

The G-L equations become 
\begin{equation}\label{eqs12v6}
\begin{array}{l}
-f_1(r)+f_1^3(r)-\frac{1}{\kappa_1 ^2}\left(\partial _r^2f_1+\frac{1}{r}\partial _rf_1\right)+\\
\frac{n^2( a-1)^2}{\kappa_1 ^2r^2}f_1 -\gamma f_2=0, 
\end{array} 
\end{equation} 

\begin{equation}\label{eqs12v7} 
\begin{array}{l} 
-\frac{\alpha _2}{\alpha _1}f_2(r)+\frac{\beta _2}{\beta _1}f_2^3(r) \\
+\frac{m_1}{m_2}\left(-\frac{1}{\kappa_1 ^2}\left(\partial _r^2f_2+\frac{1}{r}\partial _rf_2\right)+\frac{n^2(a-1)^2}{\kappa_1 ^2r^2}f_2\right) 
-\gamma f_1=0, 
\end{array} 
\end{equation} 

\begin{equation}\label{eqs12v8} 
\partial _r^2a-\frac{1}{r}\partial _ra+\left(f_1^2+\frac{m_1}{m_2}f_2^2 \right)(1-a)=0, 
\end{equation} 
and for asymptotic behavior at $ r \rightarrow \infty $ 
\begin{equation}\label{eqs12v9} 
-1+f_{10}^2-\gamma \eta =0, 
\end{equation} 
\begin{equation}\label{eq12v1} 
-\frac{\alpha _2}{\alpha _1}\eta +\frac{\beta _2}{\beta _1}\eta ^3 (1+\gamma \eta )-\gamma =0, 
\end{equation} 
$ f_{10}$ and $f_{20}$ represent the behavior of the functions $ f_{1}$ and $ f_{2}$ at infinity.  For simplicity $ f_{20} =\eta f_{10} $, where $\eta$ is a constant coefficient that relates two condensations to each other. To satisfy 
the boundary conditions at $ r \rightarrow \infty $, one can choose the functions as the follows:
\begin{equation}\label{eq12v3} 
f_1=\sqrt{1+\gamma \eta }+c_{\text{f1}}\exp \left(-\frac{r}{\sqrt{2}\xi_v}\right), 
\end{equation} 
\begin{equation}\label{eq12v4} 
f_2=\sqrt{\frac{\beta _1}{\beta _2}\left(\frac{\alpha _2}{\alpha _1}+\frac{\gamma }{\eta }\right)}+c_{\text{f2}}\exp 
\left(-\frac{r}{\sqrt{2}\xi_v}\right), 
\end{equation} 
\begin{equation}\label{eq12v5} 
a=1+c_{\text{a}}\exp \left(-\frac{r}{\lambda_v}\right). 
\end{equation} 
$\xi_{\nu}$ is equivalent to the length scale of a small fluctuation in the bulk, and it is given by the largest solution to the equation 
\begin{equation}\label{eqs12corr.} 
\left(2+3\gamma \eta-\frac{1}{2\kappa_1 ^2\xi_v ^2}\right)\left(2 \frac{\alpha _2}{\alpha _1}+3\frac{\gamma} {\eta} -\frac{m_1}{m_2}\frac{1}{2\kappa_1 ^2\xi_v ^2}\right)-\gamma ^2=0. 
\end{equation} 
 Using Eqs. (\ref{eqs12v5}), (\ref{eq12v3}) to (\ref{eq12v5}) in Eq. (\ref{eqs12v4}), the penetration depth is
\begin{equation}\label{eqs12pent} 
\lambda_v=\dfrac{1}{\sqrt{\frac{m_1}{m_2}\frac{\beta _1}{\beta _2}\left(\frac{\alpha _2}{\alpha _1}+\frac{\gamma}{\eta}\right)+(1+\gamma \eta )}} . 
\end{equation} 
 Suggesting polynomial forms for $c_{f1}$, $c_{f2}$ and $c_{a}$, the trial functions are
\begin{equation}\label{eqs12F1} 
f_1(r)=\sqrt{1+\gamma \eta }+\exp \left(-\frac{r}{\sqrt{2}\xi_v}\right)\sum _{l=0}^n\left(f_{1,l}\left.r^l\right/l!\right), 
\end{equation} 
\begin{equation}\label{eqs12F2} 
f_2(r)=\sqrt{\frac{\beta _1}{\beta _2}\left(\frac{\alpha _2}{\alpha _1}+\frac{\gamma }{\eta }\right)}+\exp \left(-\frac{r}{\sqrt{2}\xi_v}\right)\sum _{l=0}^n\left(f_{2,l}\left.r^l\right/l!\right), 
\end{equation} 
\begin{equation}\label{eqs12F3} 
a(r)=1+\exp \left(-\frac{r}{\lambda_v}\right)\sum _{l=0}^n\left(a_l\left.r^l\right/l!\right) .
\end{equation} 
\begin{figure*}[ht] 
\begin{minipage}[b]{0.5\linewidth} 
\centering 
\subfigure{$n=1$} \\
{ 
\includegraphics[width=\linewidth]{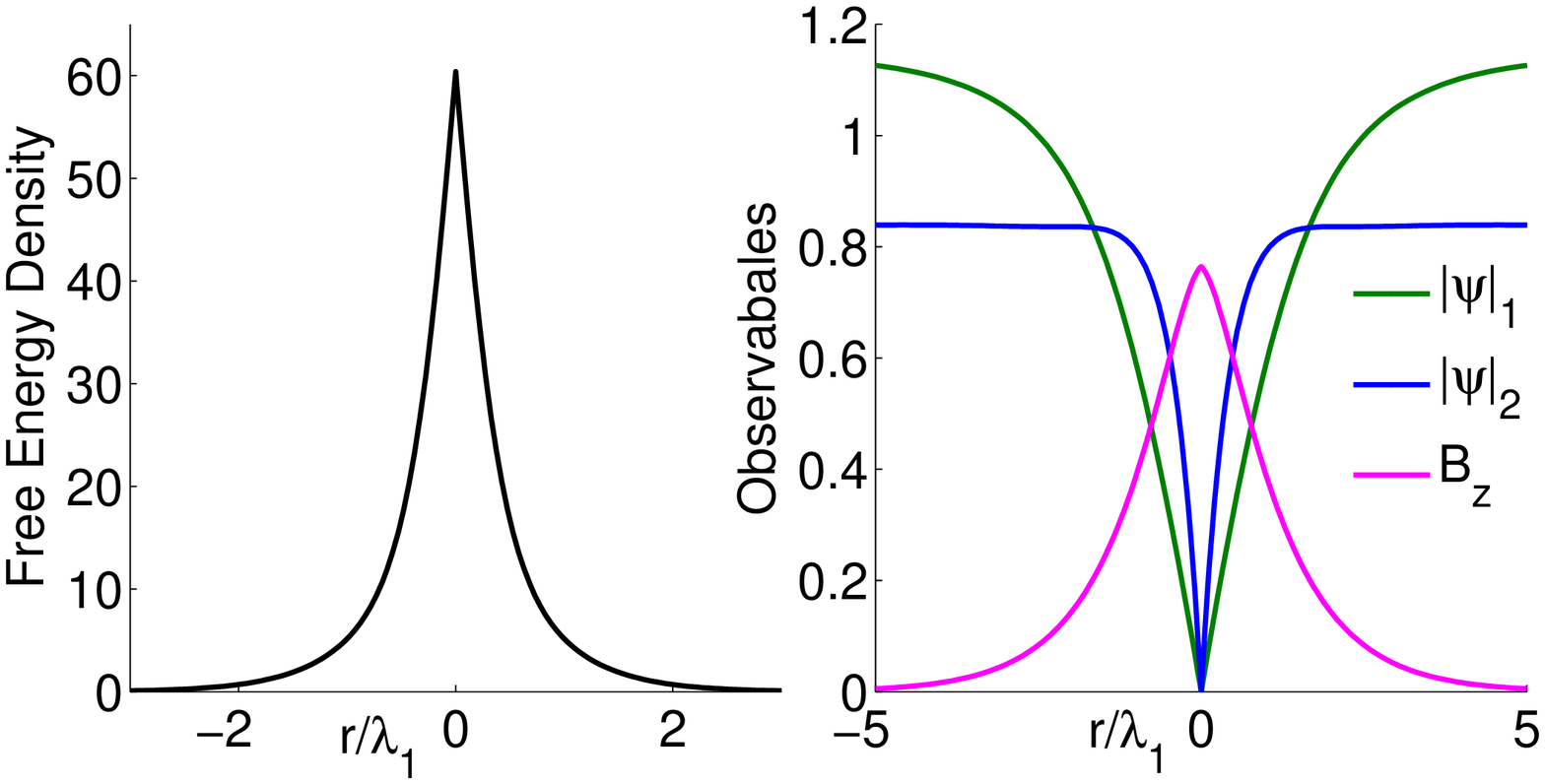} 
} 
\caption{Free energy density and condensations and magnetic field of a vortex in a type $1.5$ superconductor with 
$\xi_{1}=51nm,\xi_{2}=8nm $ and $\lambda_{1}=25nm, \lambda_{2}=30nm$ with the winding number $n=1$. The profile functions reach their asymptotic values 
at large distance.} 
\label{fig4} 
\end{minipage} 
\hspace{0.5cm} 
\begin{minipage}[b]{0.5\linewidth} 
\centering 
\subfigure{$n=2$} 
{ 
\includegraphics[width=\linewidth]{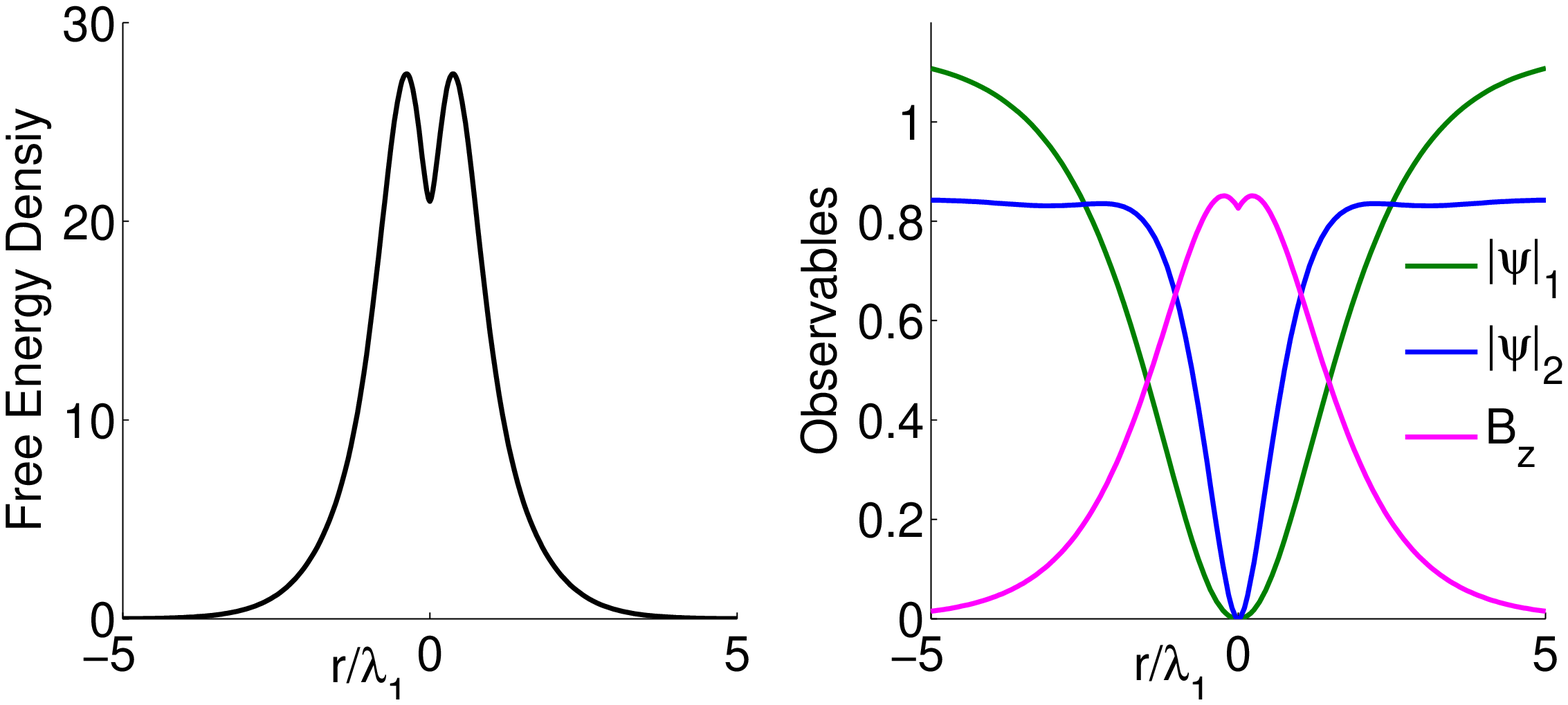} 

} 
\caption{Free energy density and condensations and magnetic field of a vortex in a type $1.5$ superconductor with $\xi_{1}=51nm,\xi_{2}=8nm $ and $\lambda_{1}=25nm,\lambda_{2}=30nm$ with the winding number $n=2$. The energies of the vortex structure are $67.4$ and $148.6$ for winding numbers $n=1$ and $n=2$, respectively, in the defined unit of the free energy. The energy of the structure in this scale is larger than type I and II superconductors.} 
\label{fig5} 
\end{minipage} 
\end{figure*} 

Boundary conditions determine some parameters and the remaining ones are variational parameters. Figure \ref{fig4} shows the condensations and 
the magnetic field for a vortex with $\xi_{1}=51nm,\xi_{2}=8nm $ and $\lambda_{1}=25nm,\lambda_{2}=30nm$ parameters. Figure \ref{fig5} shows the same fields 
for a vortex with vorticity two. 
The free energies are larger than the type I and II with the likely London penetration and condensation state. The energies of a vortex structures 
are $67.4$ and $148.6$ for winding numbers $n=1$ and $n=2$, respectively. The energy of formation of a stable vortex in these kinds of 
materials is larger than the one in type I and II.

So far we have obtained the trial  functions for the condensation states and the magnetic field. Rebbi's variational method is applied to obtain 
these solutions. We used this method for type I and II superconductors. Free energy values for $n=1$ and $n=2$ are obtained ( Fig. \ref{fig3}). We have obtained the result
that the free energy of a vortex with vorticity two is smaller than two vortices with $n=1$ for type I, so the interaction in this type of 
superconductors  is attraction. The free energy of a vortex with $n=2$ is found to be larger than the free energy of two vortices with $n=1$ in 
type II superconductors. The interaction between these vortices is repulsion.

 Now we must apply the variational method to study the interaction of type $1.5$ superconductors in which there is no monotonic interaction type for all range of distances. We must obtain the vortex profiles and magnetic field for all range of distances. To obtain trial functions of two vortices located at an arbitrary distance, Jacobs and Rebbi used conformal transformation of the complex plane $z$. $z$ is defined as $z=x+iy$. With this transformation \cite {Rebbi1979}, we have two image vortex profiles centered 
at $\pm d/2 $ in $z^\prime$ plane instead of zero in $z$ plane. For a phase change of $2\pi$ in $z^\prime$ plane there is a phase change of $4\pi$ in the $z$ plane, so this is a map of one vortex to two vortices profiles. The wave function in the complex plane can be defined as
\begin{equation}\label{eqs12F2v} 
\Psi_i \left(z,z^*\right)=\left\{\left[z^2-\left(\frac{d}{2}\right)^2\right]/\left[{z^*}^2-\left(\frac{d}{2}\right)^2\right]\right\}^{1/2}f_i\left(z,z^*\right). 
\end{equation} 

For our calculation we consider the case with equal vorticity of all the condensations. We use this projection between two polar coordinates which is different from the Jacobs coordinate system. With this projection or mathematical trick, 
one can use the trial function of one vortex to obtain the trial function of two vortices in another plane called ''$r^{\prime}-plane$". Then, it is 
possible to calculate the interaction between the vortices in this projected plane. The coordinate system of the vortices in $r^{\prime}-plane$ is defined 
by $r=r^{\prime 2}-(d/2)^2$ and $\theta^{\prime}=2\theta$. $\pm d/2$ represents the locations of the two vortices.
The trial function $f_i$ should describe not only the interaction between two separate vortices but also the solution of a giant vortex with 
vorticity two for the case when they merge to each other \cite {Rebbi1979, LinHu2011}.  
Two vortices are independent when $d \rightarrow \infty$, while at $d \sim 0$ they merge and form 
one giant vortex with vorticity two. In addition, we also need another term to describe the interaction between two vortices. Therefore, the 
trial function can be constructed as
\begin{equation}\label{12F2v2} 
\begin{array}{l} 
f_i\left(r,\theta \right)=\omega f_i^{(1)}\left(\left |r-\frac{d}{2}\right |\right)f_i^{(1)}\left(\left |r+\frac{d}{2}\right |\right)\\ 
+(1-\omega )\frac{\left |r^2-\left(\frac{d}{2}\right)^2\right |}{\left |r^2\right |}f_i^{(2)}(|r|)+\delta f_i\left(r,\theta\right), 
\end{array} 
\end{equation} 
$\delta f_i$ accounts for the interaction and $f_i^{(1)}$ and $f_i^{(2)}$ are single-vortex solutions with vorticity one and two respectively, and 
they are obtained by the method introduced for a single vortex. $\omega$ interpolates between two independent vortices 
and one giant-vortex solutions. The factor in the second term at the right-hand-side of Eq. (\ref{12F2v2}) ensures that 
the wave function vanishes at the vortex cores $r=\pm d/2$. The interaction contribution may be constructed as follows
\begin{equation}\label{eqs12F2v3} 
\begin{array}{l} 
\delta f_i \left(r,\theta \right)=\left |r^2-\left(\frac{d}{2}\right)^2\right |\frac{1}{\cosh \left(\sqrt{2}\kappa_1 |r|\right)}\\
\sum _{l=0}^n\sum _{j=0}^l f_{i,lj}\frac{\left |r|^{2l}\right.}{2}\left[\left(e^{(2I\theta )}\right)^j+\left(e^{(-2I\theta)}\right)^j\right]. 
\end{array} 
\end{equation} 

The first factor is to make sure that the wave function vanishes at the vortex cores, and the second factor accounts for the fact that 
the interaction vanishes when $r \rightarrow \infty$. $I$ in the exponentials represents $i=\sqrt{-1}$ which is typed in capital form to avoid any 
confusion with the ''i" in the summation. When we put two vortices in a plane, the circular symmetry would be lost. Only a reflection symmetry with respect to the plane
would remain. The polynomial in the above equation preserves such a reflection symmetry.

The same procedure which is applied to $f_i$ for constructing $\psi_i$ applies to $\mathbf{A}$: 
\begin{equation}\label{eqs3n5} 
\begin{array}{l} 
\mathbf{A}=\omega \left[\frac{ 1}{\kappa_1 \left |r-d/2\right |}a^{(1)}\left(\left |r-\frac{d}{2}\right |\right)+\frac{ 1}{\kappa_1 \left |r+d/2\right |}a^{(1)}\left(\left |r+\frac{d}{2}\right |\right)\right]\\ 
+\frac{2 }{\kappa_1 r}(1-\omega )a^{(2)}(|r|)+\delta a\left(r,\theta \right), 
\end{array} 
\end{equation} 

where $a^{(1)}$ and $a^{(2)}$ are functions of the single-vortex solutions 
with vorticities one and two. The asymptotic behavior of vortices implies the interaction contribution and it has the following form:
 \begin{equation}\label{eqs3n6} 
\delta a\left(r,\theta\right)=\frac{1}{\cosh (|r|)}\left[ra_1\left(r,\theta\right)+r a_2\left(r,\theta\right)\right], 
\end{equation}
with 
\begin{equation}\label{eqs3n7} 
a_k\left(r,\theta\right)=\sum _{i=0}^n\sum _{j=0}^ia_{{k,ij}}\frac{\left |r|^{2i}\right.}{2}\left[\left(e^{(2I\theta)}\right)^j+\left(e^{(-2I\theta)}\right)^j\right], 
\end{equation}
where $k=1, 2$. $f_{i,lj}$ and $a_{k,ij}$ are new variational parameters which must be obtained numerically. We consider the variational parameters up to the coefficients of $|r|^{6}$ in our calculations.
 
\begin{figure}
\hspace{-1cm}  
\begin{minipage}[b]{0.5\linewidth} 
\centering 
\subfigure{$ d=1$}\\ 
{ 
\includegraphics[width=1\linewidth]{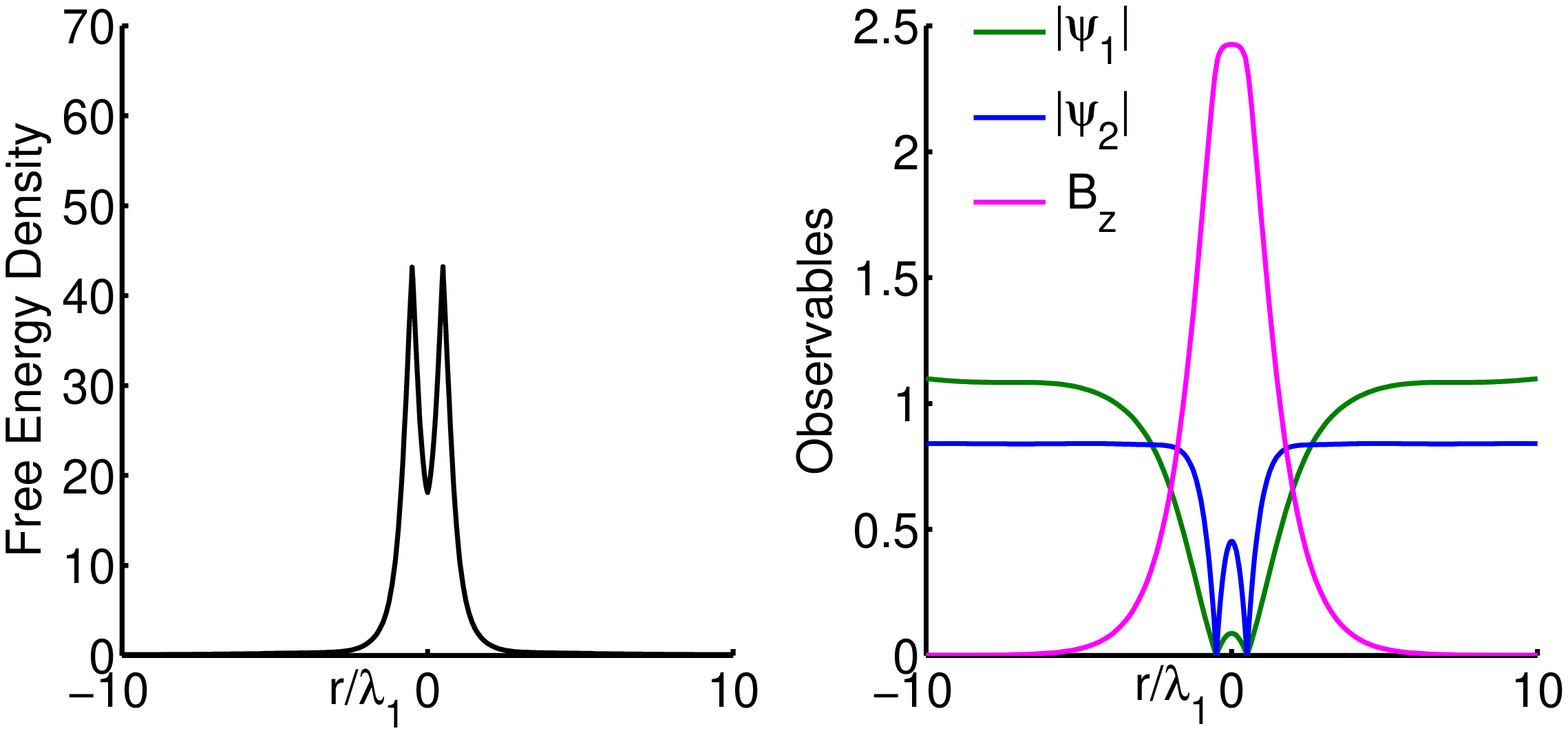} 
\label{secondOtwoco1} 
} 
\\ 
\subfigure{$d=2$}\\ 
{
\includegraphics[width=1\linewidth]{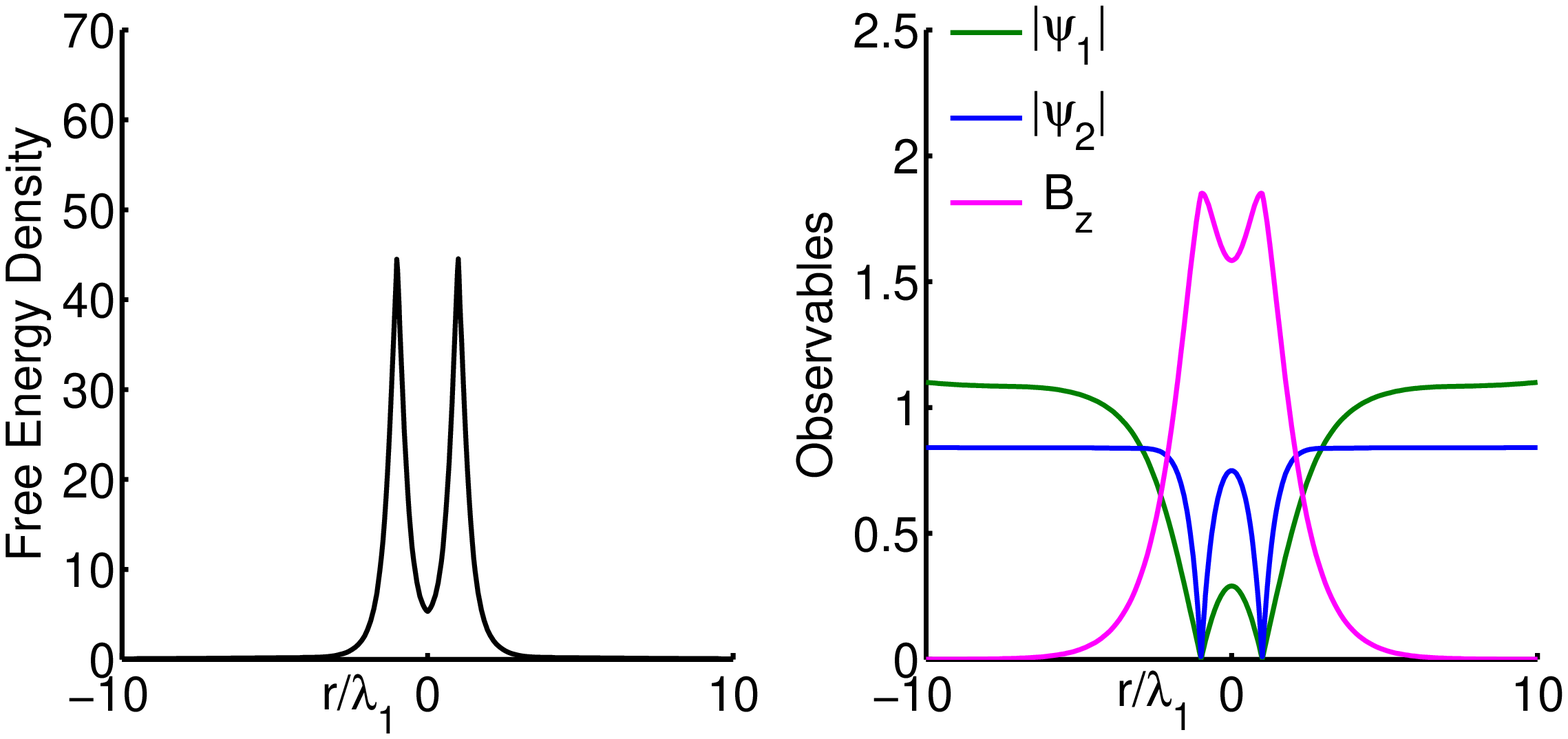} 
\label{secondOtwoco22} 
} 
\\ 
\subfigure{$d=3$}\\ 
{
\includegraphics[width=1\linewidth]{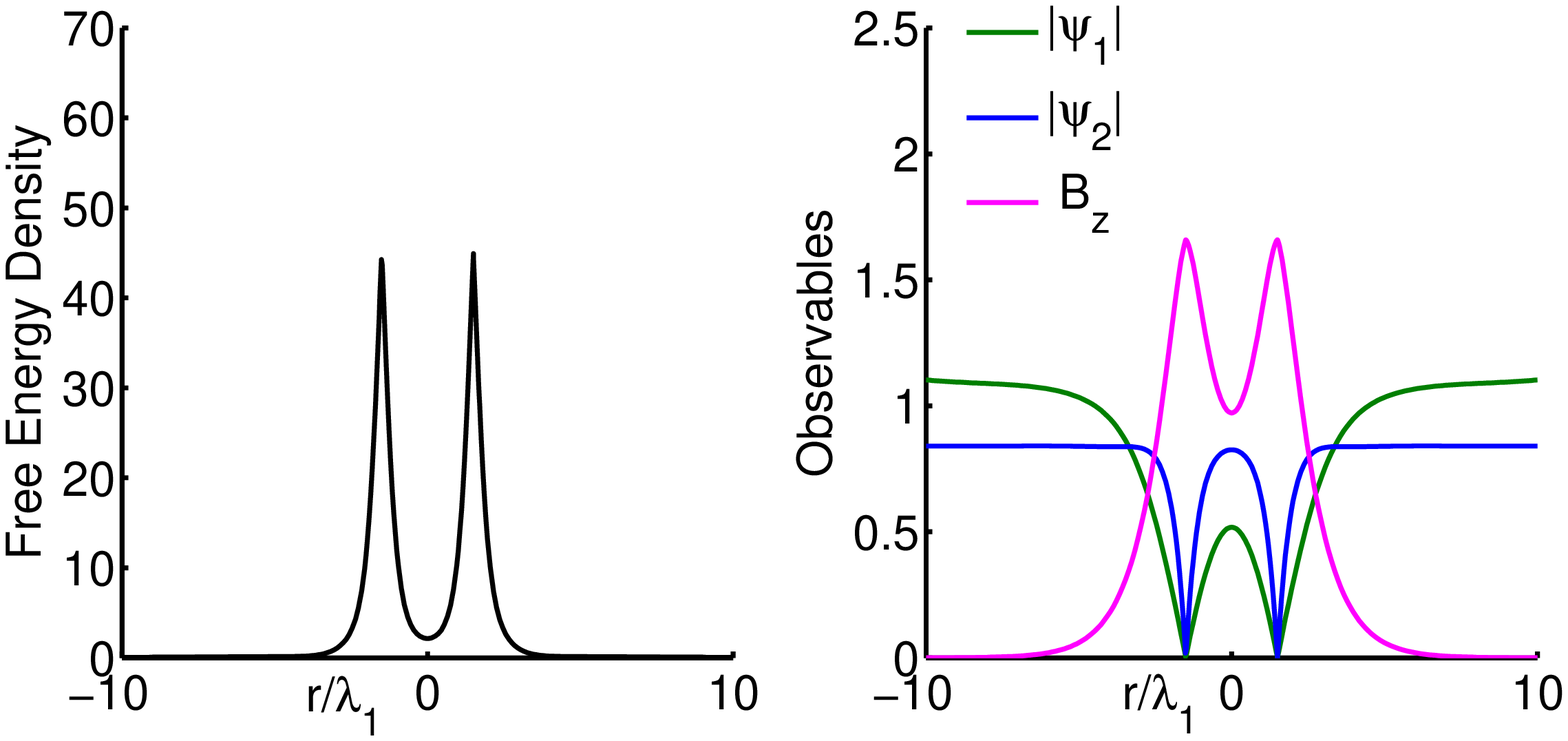} 
\label{secondOtwoco3} 
} 
\\ 
\subfigure{$d=4$}\\ 
{
\includegraphics[width=1\linewidth]{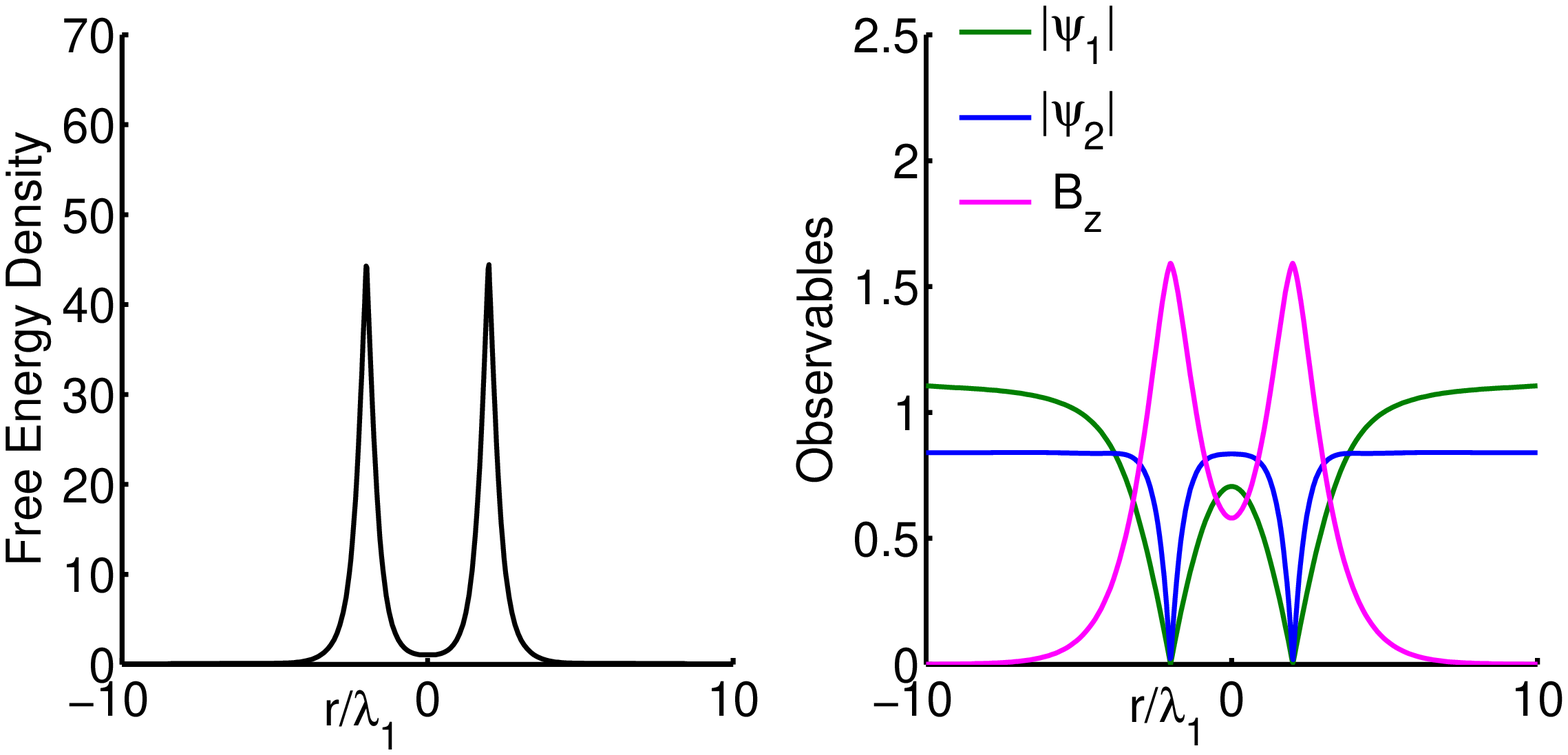} 
\label{secondOtwoco4} 
} 
 
\end{minipage} 
\begin{minipage}[b]{0.5\linewidth} 
\centering 
\subfigure{$d=5$}\\ 
{
\includegraphics[width=1\linewidth]{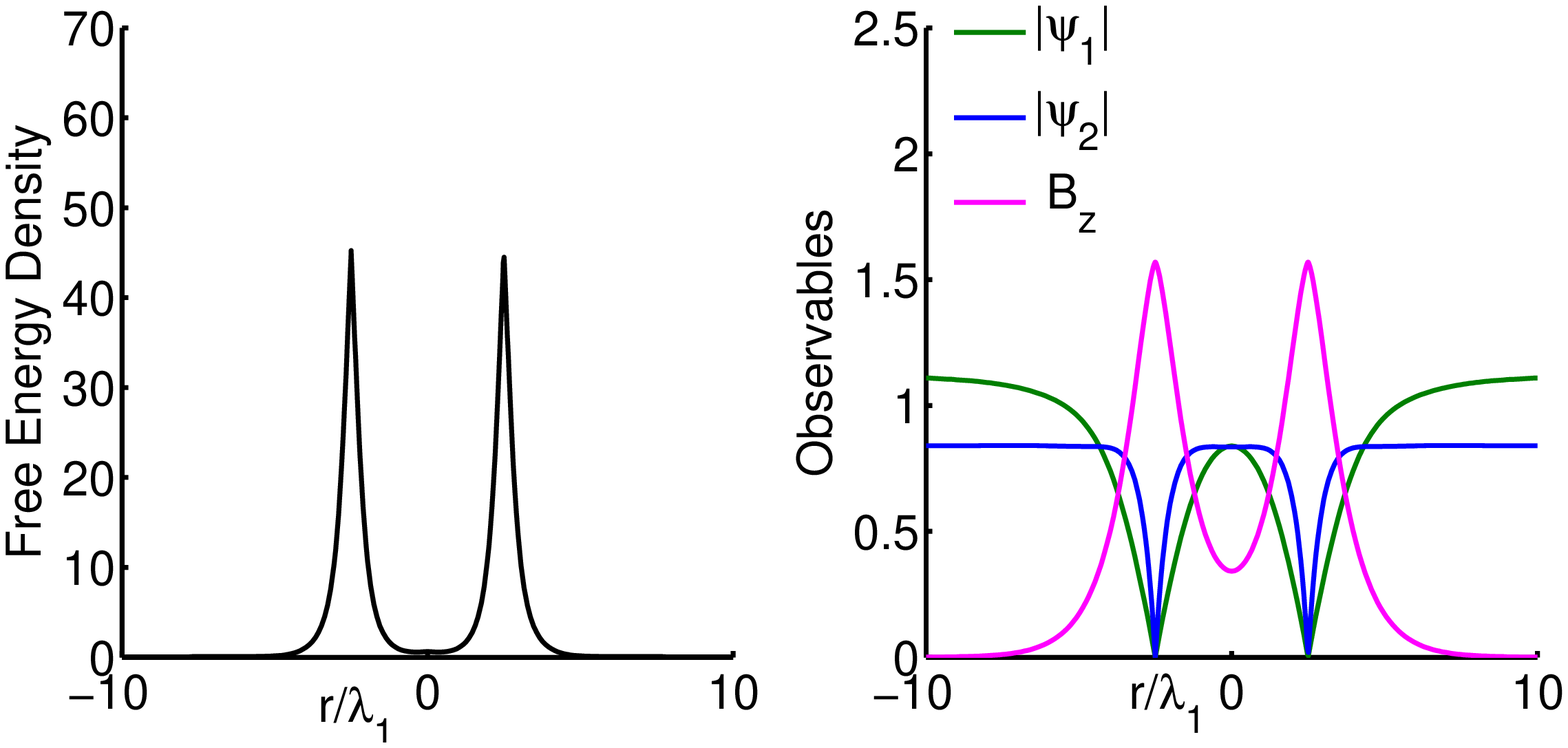} 
\label{secondOtwoco5} 
}
\\
\subfigure{$d=6$}\\ 
{ 
\includegraphics[width=1\linewidth]{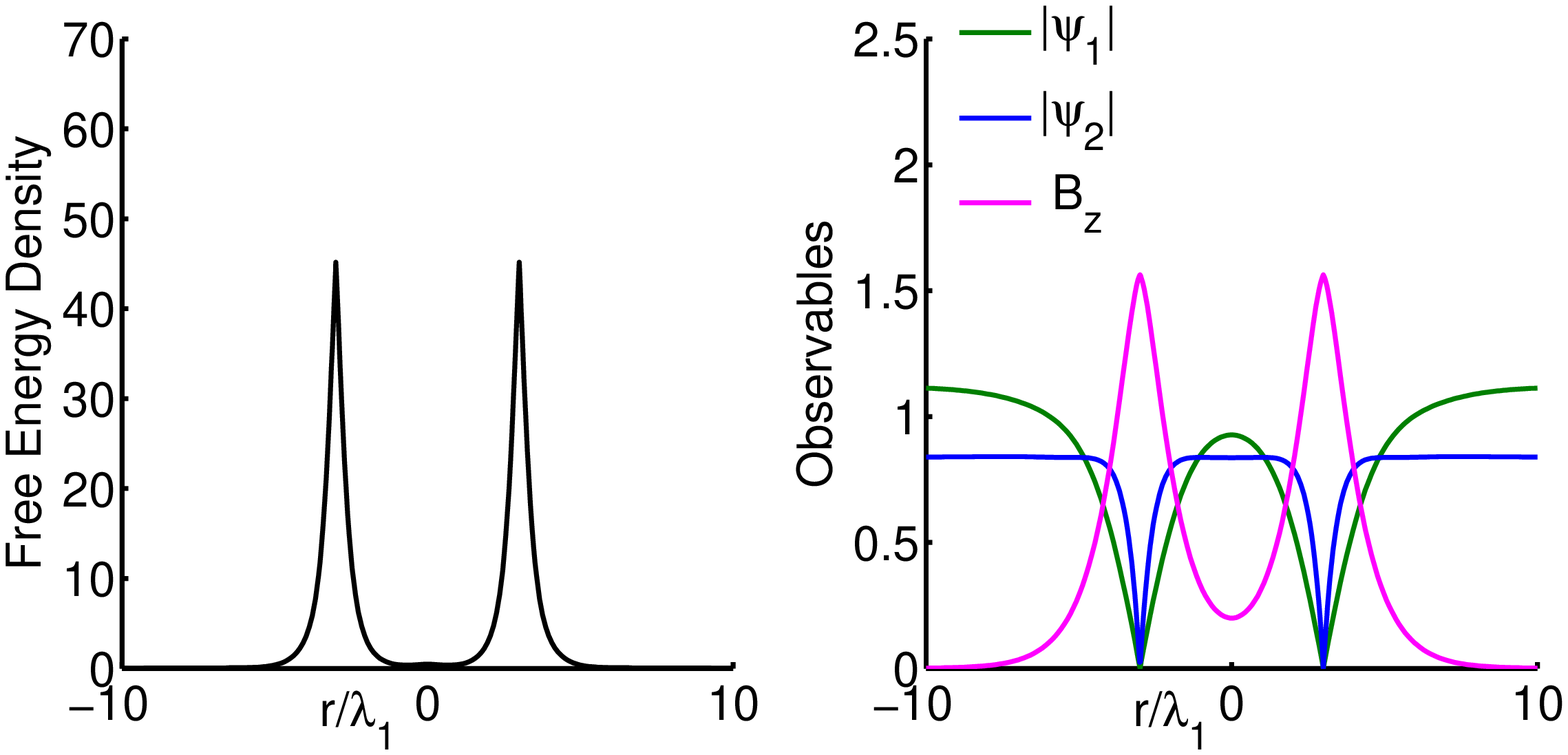} 
\label{secondOtwoco6} 
} 
\\ 
\subfigure{$d=7$}\\ 
{ 
\includegraphics[width=1\linewidth]{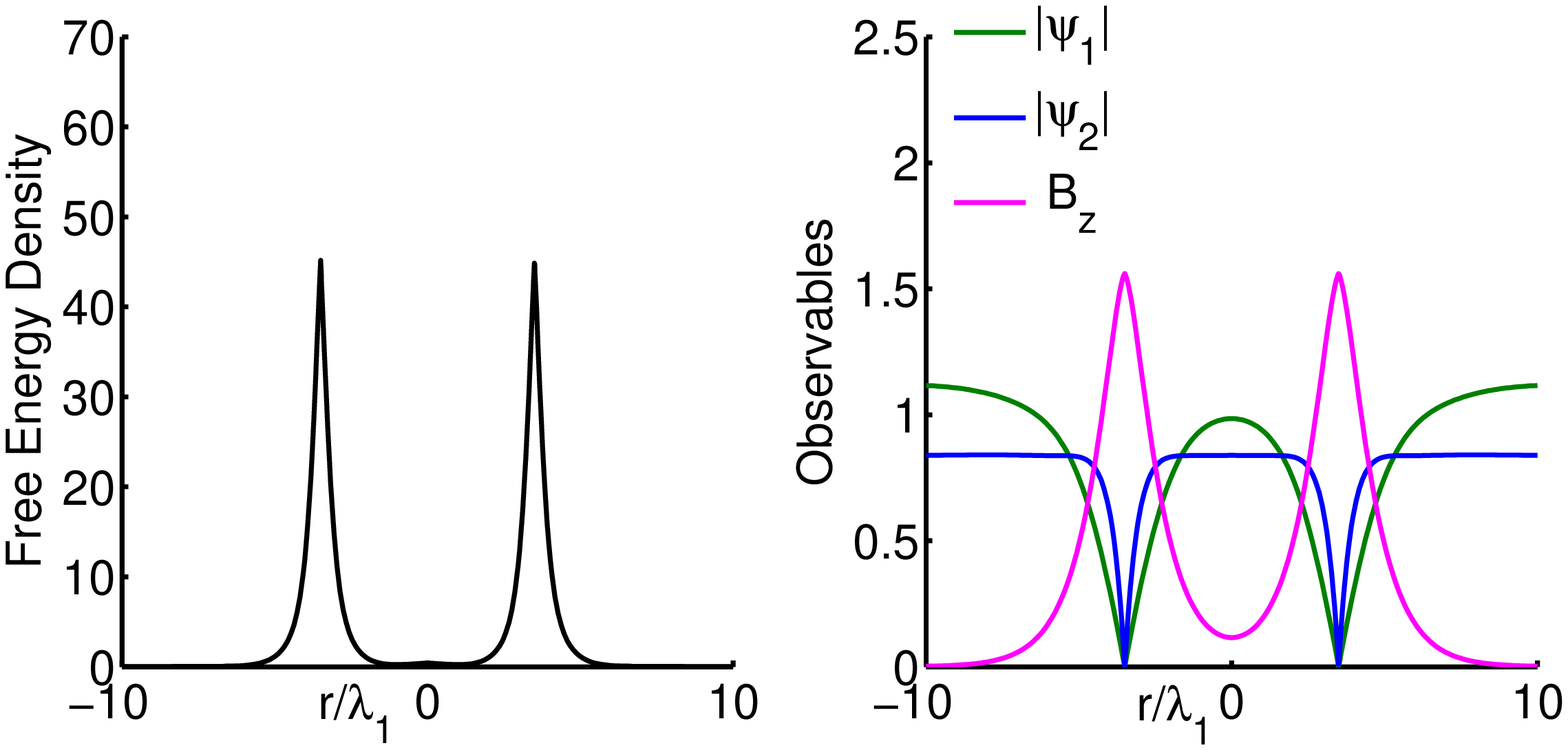} 
\label{secondOtwoco7} 
} 
\\ 
\subfigure{$d=8$}\\ 
{
\includegraphics[width=1\linewidth]{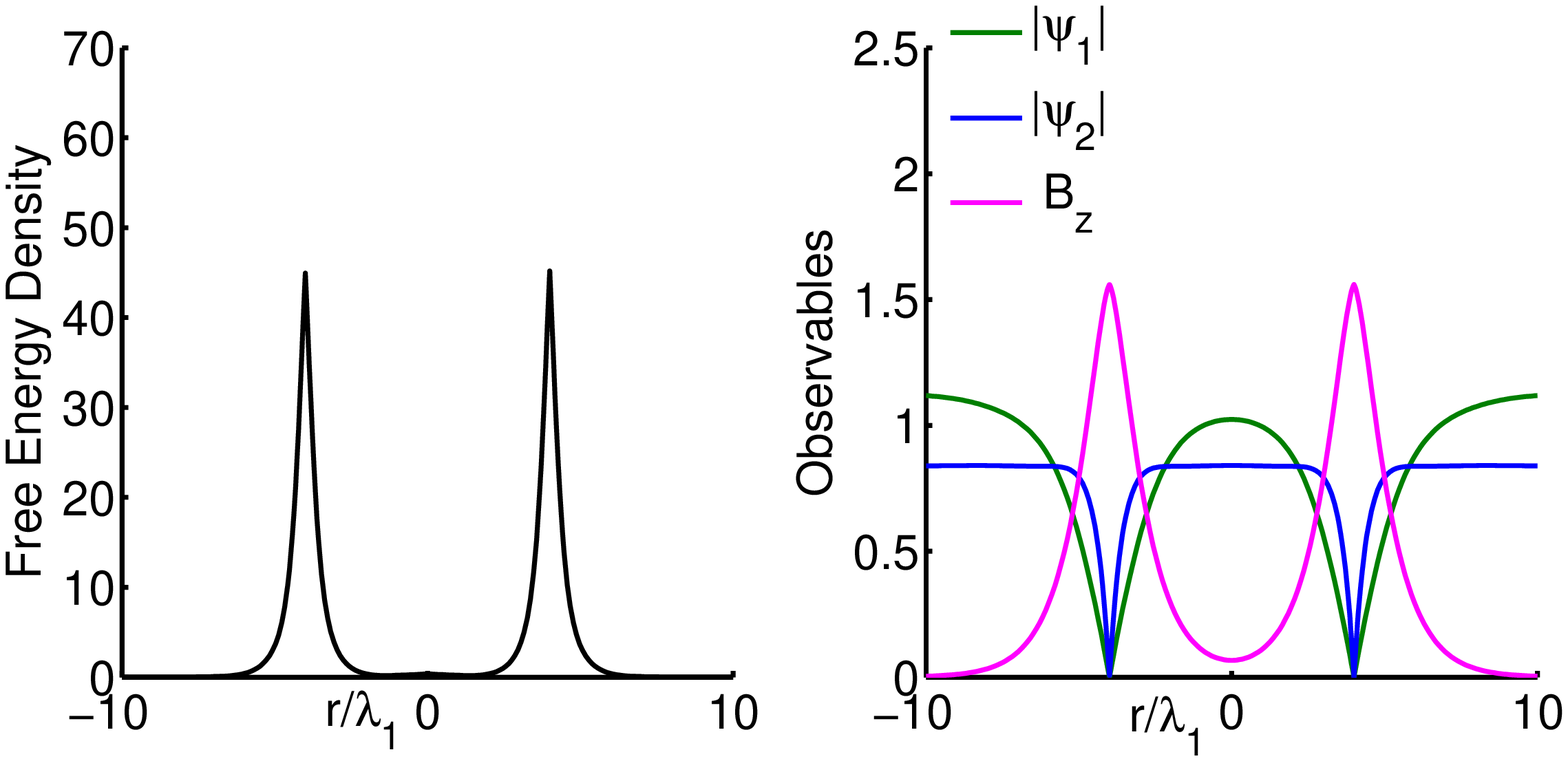} 
\label{secondOtwoco8} 
} 
\end{minipage} 

\caption{Free energy density, condensation states, and magnetic field profiles cross sections in a plane when two vortices of type $1.5$ are located at different distances. 
As the separation between the vortices increases the magnetic field profile function decreases  between the vortices. When the vortices are far 
from each other they are like two separate vortices with no interaction.} 
\label{En2} 
\end{figure} 
\begin{figure}
\centering 
\includegraphics[width=0.5\linewidth]{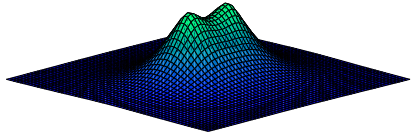} 
\caption{(color online) A three dimensional plot of the magnetic field of two vortices of type $1.5$ at distance $d=2$. No circular symmetry is present any more. 
Only a reflection symmetry with respect to the plane at the middle distance between the vortices is present} 
\label{3D} 
\end{figure} 

\begin{figure}
\centering 
\includegraphics[width=0.5\linewidth]{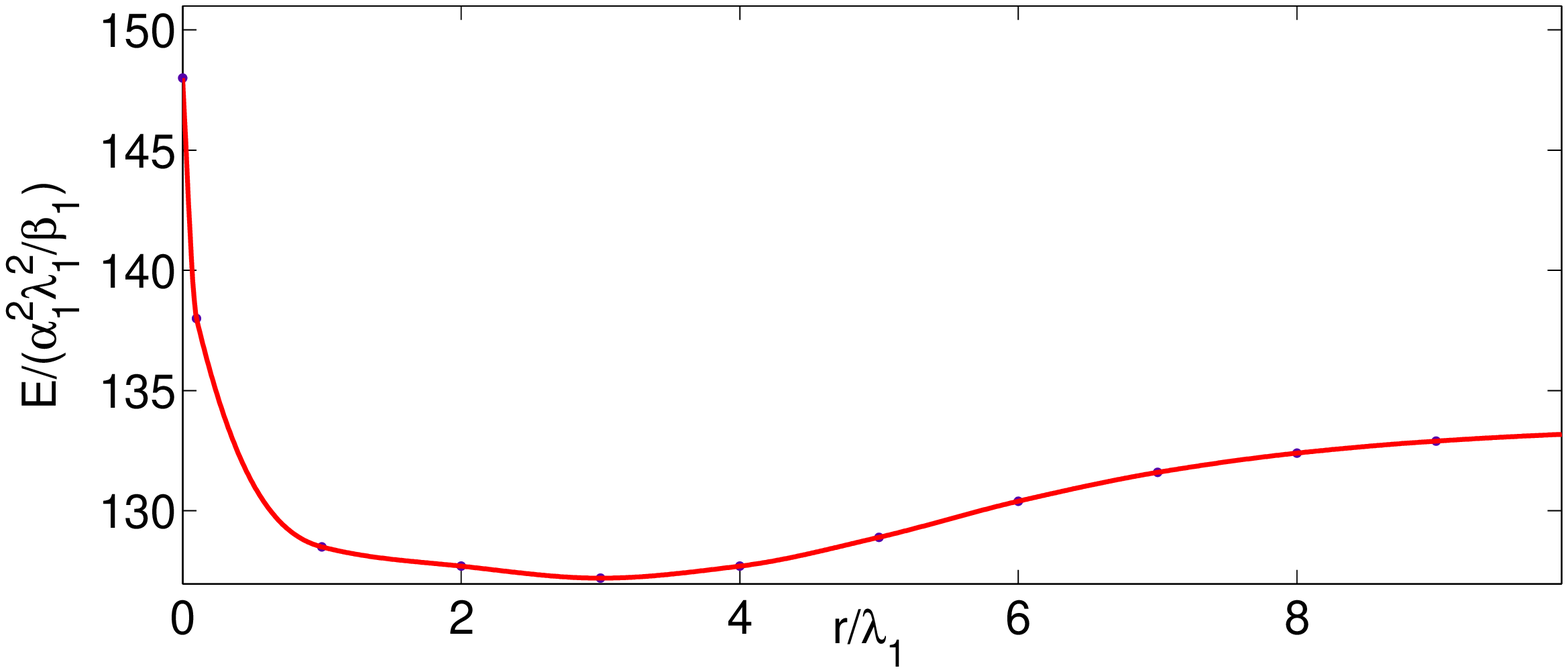} 
\caption{Energy of two vortices of type $1.5$, located at different distances. This shows a stability point at $2.7$. At larger distance than $2.7$ this 
energy increases, so at these separations the interaction is attraction. At smaller distances than $2.7$, the energy increases and leads to 
repulsion between the vortices} 
\label{ED2} 
\end{figure} 

Figure (\ref{En2}) shows the condensation states, magnetic field and free energy density between two type $1.5$ vortices at different 
distances. Figure (\ref{fig5}) and (\ref{En2}) shows that by increasing $d$ the distribution of the magnetic field changes such that for large 
$d$, each vortex has its own magnetic field, almost independently. However as $d \rightarrow 0$, the magnetic field is distributed along the 
vortex with vorticity two, as expected.
In figure (\ref{3D}) we show a three dimensional magnetic field of two vortices at the distance $d=2$. Only the so called reflection symmetry remains. 
Figure (\ref{ED2}) shows the interaction energy versus distance between two vortices. As the distance between vortices decreases the energy 
decreases up to distance $2.7$, so the interaction between two vortices in this range of distances, is attraction. The energy increases from 
distance $2.7$ to zero, so the interaction  is repulsion. Our results agree the results obtained in Ref. \cite{LinHu2011}. We also obtain the same 
stability point by choosing  the same penetration depth and correlation length but  with other different parameters such as $\gamma$. In addition, we 
use polar coordinate instead of a Cartesian coordinate. The polar coordinate simplifies  calculations when we have only a vortex in the plane with circular symmetry. 
When two vortices are imposed in a plane,  this circular symmetry is lost and only a reflection symmetry with respect to the plane at the middle distance from center of vortices survives. 
By losing the circular symmetry, the $\theta$ dependence of functions is included again.

\section{Interaction between Vortices with Three Condensation States} 
What about the situation with three condensation states? The idea of vortex with three condensation states can be used to describe the iron based superconductors. 
Also, Babaev and Weston have recently studied the possibility of existence of more than three condensation states from theoretical point of 
view \cite{Babaev and weston}. We use the method of previous section for a case with three condensationOnly an introduceds. For simplicity, we study the cases 
for which the interband scattering couplings are equal. 
The equations of motions are obtained by using (\ref{eqsFd1}) for G-L free energy for three states
\begin{equation}\label{eqs3v1} 
-\Psi _1+\left |\Psi _1\right |{}^2\Psi _1+\left(\frac{1}{i \kappa _1}\nabla -\textbf{A}\right){}^2\Psi _1-\gamma \Psi _2-\gamma \Psi _3=0, 
\end{equation} 
\begin{equation}\label{eqs3v2} 
\begin{array}{l}
-\frac{\alpha _2}{\alpha _1}\Psi _2+\frac{\beta _2}{\beta _1}\left |\Psi _2\right |{}^2\Psi _2+\frac{m_1}{m_2}\left(\frac{1}{i \kappa _1}\nabla -\textbf{A}\right){}^2\Psi_2\\
-\gamma \Psi _1-\gamma \Psi _3=0, 
\end{array}
\end{equation} 
\begin{equation}\label{eqs3v3}
\begin{array}{l}
-\frac{\alpha _3}{\alpha _1}\Psi _3+\frac{\beta _3}{\beta _1}\left |\Psi _3\right |{}^2\Psi _3+\frac{m_1}{m_3}\left(\frac{1}{i \kappa _1}\nabla -\textbf{A}\right){}^2\Psi_2-\\
\gamma \Psi _1-\gamma \Psi _2=0, 
\end{array} 
\end{equation} 
\begin{equation}\label{eqs3v4} 
\begin{array}{l} 
\nabla \times \nabla \times \textbf{A}=\frac{1}{2i \kappa _1}(\Psi _1^*\nabla \Psi _1-\Psi _1\nabla \Psi _1^*)-\left |\Psi _1\right |{}^2\textbf{A}\\ 
+\frac{m_1}{m_2}\left(\frac{1}{2i \kappa _1}(\Psi _2^*\nabla \Psi _2-\Psi _2\nabla \Psi _2^*)-\left |\Psi _2\right |{}^2\textbf{A}\right)+\\
\frac{m_1}{m_3}\left(\frac{1}{2i \kappa _1}(\Psi _3^*\nabla \Psi _3-\Psi _3\nabla \Psi _3^*)-\left |\Psi _3\right |{}^2\textbf{A}\right). 
\end{array} 
\end{equation} 
As mentioned above, $\gamma_1=\gamma_2=\gamma_3$ which means that the strength of all interband couplings are equal. 
Equation (\ref{eqs3v4}) describes the screening of the magnetic field by the superconducting condensates. Again, using the London 
approximation, the effective London penetration depth for three-band superconductors is 
\begin{equation}\label{eqs3vpen} 
\lambda_v =1\left/\sqrt{\left |\Psi _{10}\right |{}^2+\frac{m_1}{m_2}\left |\Psi _{20}\right |{}^2+\frac{m_1}{m_3}\left |\Psi _{30}\right |{}^2}\right. , 
\end{equation} 
where $\Psi_{i0}$ is the bulk value of the $i$th superconducting condensate. 
Since all the response of three-band superconductors to the magnetic fields is described by a single length scale $\lambda$, all condensates couple 
to the same gauge field. 
The interband coupling changes the bulk value, and it modifies the corresponding penetration depth. 
Again, we use the so-called ansatz (\ref{eqsEM3}) and obtain the equations 
\begin{equation}\label{eqs3v5} 
\begin{array}{l}
-f_1(r)+f_1^3(r)-\frac{1}{\kappa_1 ^2}\left(\partial _r^2f_1+\frac{1}{r}\partial _rf_1\right)+\frac{n^2( a-1)^2}{\kappa_1 ^2r^2}f_1\\
-\gamma f_2-\gamma f_3=0, 
\end{array}
\end{equation} 
\begin{equation}\label{eqs3v6} 
\begin{array}{l} 
-\frac{\alpha _2}{\alpha _1}f_2(r)+\frac{\beta _2}{\beta _1}f_2^3(r) +\frac{m_1}{m_2}\\
\left(-\frac{1}{\kappa_1 ^2}\left(\partial _r^2f_2+\frac{1}{r}\partial _rf_2\right)+\frac{n^2( a-1)^2}{\kappa_1 ^2r^2}f_2\right)-\gamma f_1-\gamma f_3=0, 
\end{array} 
\end{equation} 
\begin{equation}\label{eqs3v7} 
\begin{array}{l} 
-\frac{\alpha _3}{\alpha _1}f_3(r)+\frac{\beta _3}{\beta _1}f_3^3(r) +\\
\frac{m_1}{m_3}\left(-\frac{1}{\kappa_1 ^2}\left(\partial _r^2f_2+\frac{1}{r}\partial _rf_3\right)+\frac{n^2( a-1)^2}{\kappa_1 ^2r^2}f_3\right)-\gamma f_1-\gamma f_2=0, 
\end{array} 
\end{equation} 
\begin{equation}\label{eq3v7} 
\partial _r^2a-\frac{1}{r}\partial _ra+\left(f_1^2+\frac{m_1}{m_2}f_2^2+\frac{m_1}{m_3}f_3^2\right)(1-a)=0. 
\end{equation} 
In the limit when $r\rightarrow\infty$, the wave functions are defined by the 
bulk values $f_{10}$ and $f_{20}$ and $f_{30}$. Defining $f_{20}=\eta f_{10}$ 
with $\eta>0$ and $f_{30}=\eta' f_{10}$ with $\eta'>0$, we have the equations for $f_{10}$ and $\eta$ and $\eta'$ 
\begin{equation}\label{eq3v8} 
-1+f_{10}^2-\gamma \eta-\gamma \eta' =0, 
\end{equation} 
\begin{equation}\label{eq3v9} 
-\frac{\alpha _2}{\alpha _1}\eta +\frac{\beta _2}{\beta _1}\eta ^3 \left(1+\gamma \eta+\gamma \eta' \right)-2\gamma =0, 
\end{equation} 
\begin{equation}\label{eq3v10} 
-\frac{\alpha _3}{\alpha _1}\eta' +\frac{\beta _3}{\beta _1}\eta ^{'3} \left(1+\gamma \eta+\gamma \eta' \right)-2\gamma =0. 
\end{equation} 
The radial variation of the wave functions and  the vector potential in the asymptotic region for $r\rightarrow\infty$ is found and is given by 
\begin{equation}\label{eqs3vF1} 
f_1=\sqrt{1+\gamma \eta+\gamma \eta' }+c_{\text{f1}}\exp \left(-\frac{r}{\sqrt{2}\xi_v}\right), 
\end{equation} 
\begin{equation}\label{eqs3vF2} 
f_2=\sqrt{\frac{\beta _1}{\beta _2}\left(\frac{\alpha _2}{\alpha _1}+\frac{\gamma+\gamma }{\eta+\eta' }\right)}+c_{\text{f2}}\exp 
\left(-\frac{r}{\sqrt{2}\xi_v}\right), 
\end{equation} 

\begin{equation}\label{eqs3vF3} 
f_3=\sqrt{\frac{\beta _1}{\beta _3}\left(\frac{\alpha _3}{\alpha _1}+\frac{\gamma+\gamma }{\eta+\eta' }\right)}+c_{\text{f3}}\exp 
\left(-\frac{r}{\sqrt{2}\xi_v}\right), 
\end{equation} 
\begin{equation}\label{eqs3vF4} 
a=1+c_{\text{a}}\exp \left(-\frac{r}{\lambda_v}\right). 
\end{equation} 
At large distances, there is only one length scale for the three condensates, called the penetration depth $\lambda_v$. It can be obtained 
straightforwardly from Eqs. (\ref{eqs3vF1}), (\ref{eqs3vF2}), (\ref{eqs3vF3}) and (\ref{eqs3vpen}) 
\begin{equation}\label{eqs3vpenvvv} 
\begin{array}{l}
\lambda_v=\\
1\left/\sqrt{\sum \limits_{p=2,3}\frac{m_1}{m_p}\frac{\beta _1}{\beta _p}\left(\frac{\alpha _p}{\alpha _1}+\frac{\gamma+\gamma }{\eta+\eta' }\right)+
(1+\gamma \eta+\gamma \eta' )}\right . 
\end{array}
\end{equation} 
To calculate the correlation length, we substitute the asymptotic limits of Eqs.(\ref{eqs3vF1}) to (\ref{eqs3vF4}) into Eqs.(\ref{eqs3v4}) to (\ref{eqs3v7}) and linearize the equations by considering only the linear parts of the terms. The following equation is obtained as a result of these procedures:
\begin{equation}\label{eqs3vcorr} 
\begin{array}{l}
\left(2+3\gamma \eta-\frac{1}{2\kappa_1 ^2\xi_v ^2}\right)\left(2 \frac{\alpha _2}{\alpha _1}+3\frac{\gamma} {\eta} -\frac{m_1}{m_2}\frac{1}{2\kappa_1 ^2\xi_v ^2}\right)\\
\left(2 \frac{\alpha _3}{\alpha _1}+3\frac{\gamma'} {\eta'} -\frac{m_1}{m_3}\frac{1}{2\kappa_1 ^2\xi_v ^2}\right)-\gamma ^3=0. 
\end{array}
\end{equation} 
 $\xi_v$ is equivalent to the length scale of the small fluctuations in the bulk and is given by the largest solution to the equation.
$\xi_v$ is an effective length which is in fact the correlation in a system with interband coupling. With this new definition, the trial functions
 become 
\begin{equation}\label{eqs3vtf} 
f_1(r)=\sqrt{1+2\gamma \eta }+\exp \left(-\frac{r}{\sqrt{2}\xi_v}\right)\sum _{l=0}^n\left(f_{1,l}\left.r^l\right/l!\right), 
\end{equation} 
\begin{equation}\label{eqs3vtff} 
f_2(r)=\sqrt{\frac{\beta _1}{\beta _2}\left(\frac{\alpha _2}{\alpha _1}+\frac{\gamma }{\eta }\right)}+\exp \left(-\frac{r}{\sqrt{2}\xi_v}\right)\sum _{l=0}^n\left(f_{2,l}\left.r^l\right/l!\right), 
\end{equation} 
\begin{equation}\label{eqs3vtfff} 
f_3(r)=\sqrt{\frac{\beta _1}{\beta _3}\left(\frac{\alpha _3}{\alpha _1}+\frac{\gamma }{\eta }\right)}+\exp \left(-\frac{r}{\sqrt{2}\xi_v}\right)\sum _{l=0}^n\left(f_{3,l}\left.r^l\right/l!\right), 
\end{equation}
\begin{equation}\label{eqs3vta} 
a(r)=1+\exp \left(-\frac{r}{\lambda_v}\right)\sum _{l=0}^n\left(a_l\left.r^l\right/l!\right) , 
\end{equation} 
where $f_{1,l}$, $f_{2,l}$ , $f_{3,l}$ and $a_l$ are variational parameters. 
Following the procedure of the previous sections, we obtain the 
variational coefficients, from which we can obtain the vortex 
solution. We truncate the higher-order corrections of the trial functions at $n=6$ and find the solution of a single vortex 
with vorticity one and two. The penetration depths and correlation lengths we consider for our calculation are $\xi_{1}=51nm,\xi_{2}=8nm,\xi_{3}=25nm $ and $\lambda_{1}=25nm,\lambda_{2}=30nm,\lambda_{3}=51nm$. We take $\gamma=\gamma'=0.4 >0$ and also $\eta=\eta'=0.5$ for our calculation. 
\begin{figure*}[ht] 
\begin{minipage}[b]{0.5\linewidth} 
\centering 
\subfigure{\hspace{1cm} $n=1$} 
\includegraphics[width=1\linewidth]{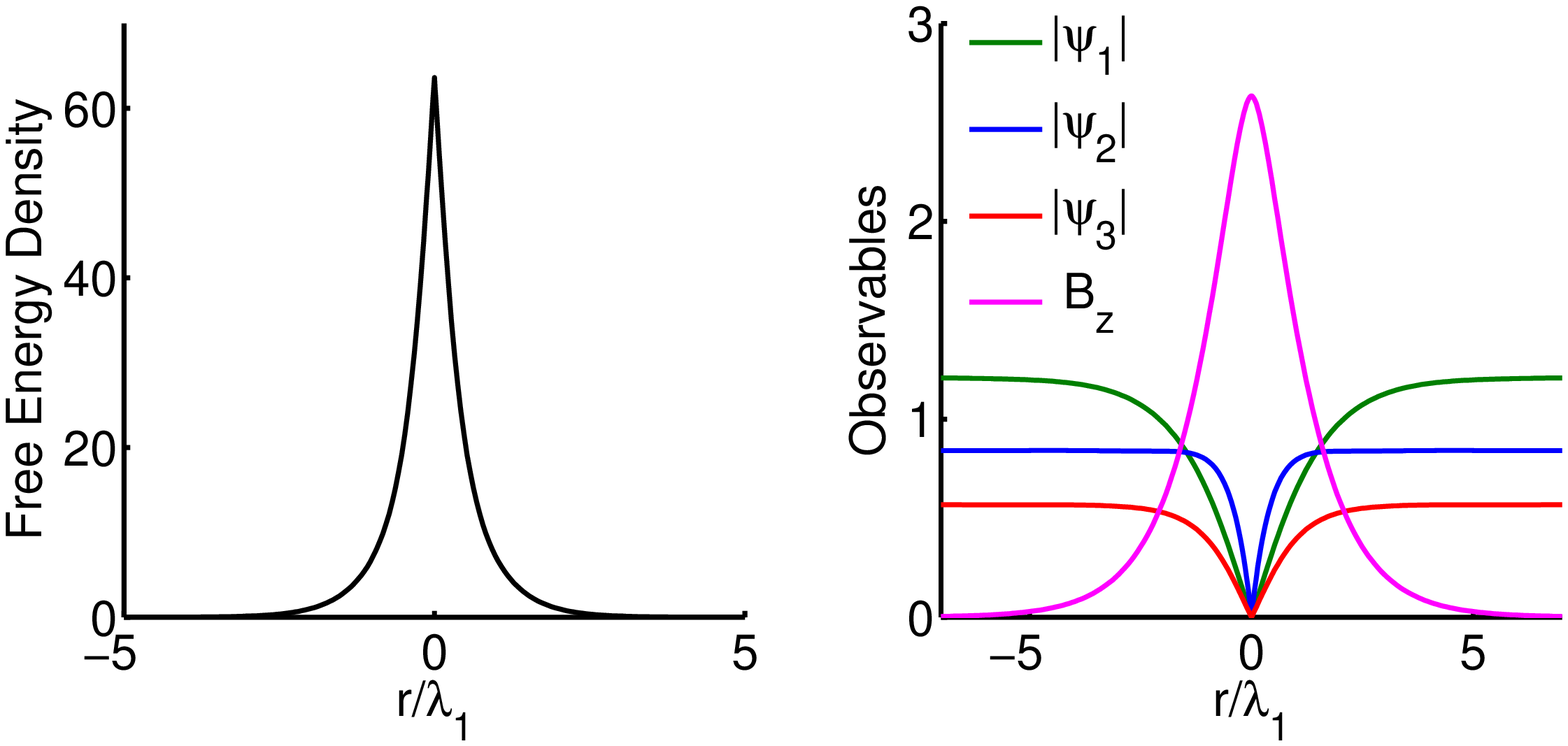} 
\end{minipage} 
\begin{minipage}[b]{0.5\linewidth} 
\centering 
{ 
\subfigure{\hspace{1cm} $n=2$} 
\includegraphics[width=1\linewidth]{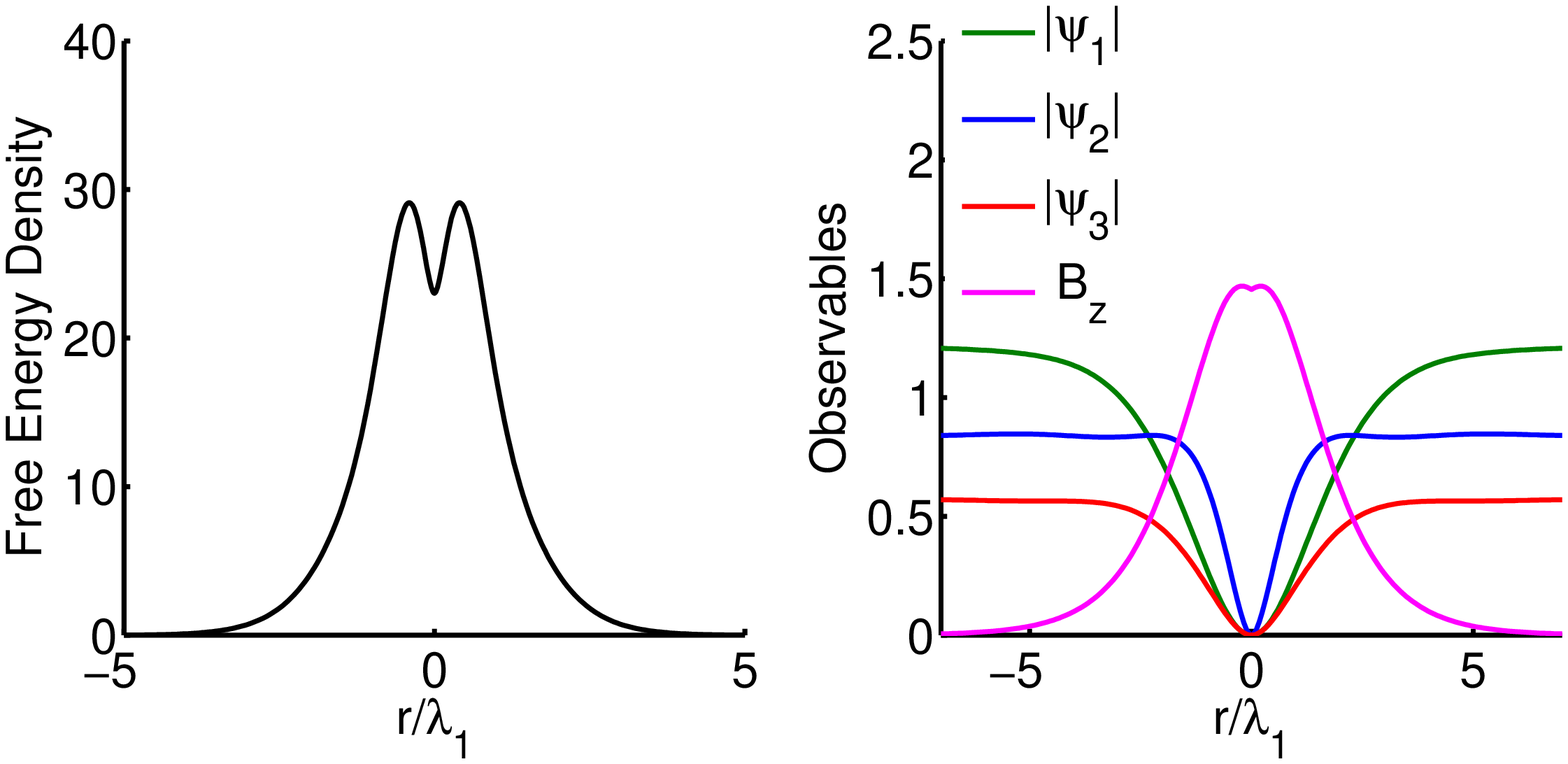} 
} 
\end{minipage}  
\caption{Free energy density and condensations and magnetic field of a vortex with three condensation states with 
$\xi_{1}=51nm,\xi_{2}=8nm,\xi_{3}=25nm $ and $\lambda_{1}=25nm,\lambda_{2}=30nm,\lambda_{3}=51nm$, using the variational method but a polar coordinate is used. The energy of vortex is $85.2$ for $n=1$ and $189.5$ for $n=2$ in the defined dimension unit. The energy of structure in this scale is larger than type I and II 
superconductors.} 
\label{OVECR}
\end{figure*}  
 
The energy of the vortex is $85.2$ for $n=1$ and $189.5$ for $n=2$. We can see the role of increasing the number of condensations in increasing 
the energy of formation of a vortex in these materials. The phenomenon has been observed when we had two condensations compared with 
the case when we had one condensation.  
\begin{figure}
\hspace{-1cm} 
\begin{minipage}[b]{0.5\linewidth} 
\centering 
\subfigure{$d=1$}\\ 
{
\includegraphics[width=1\linewidth]{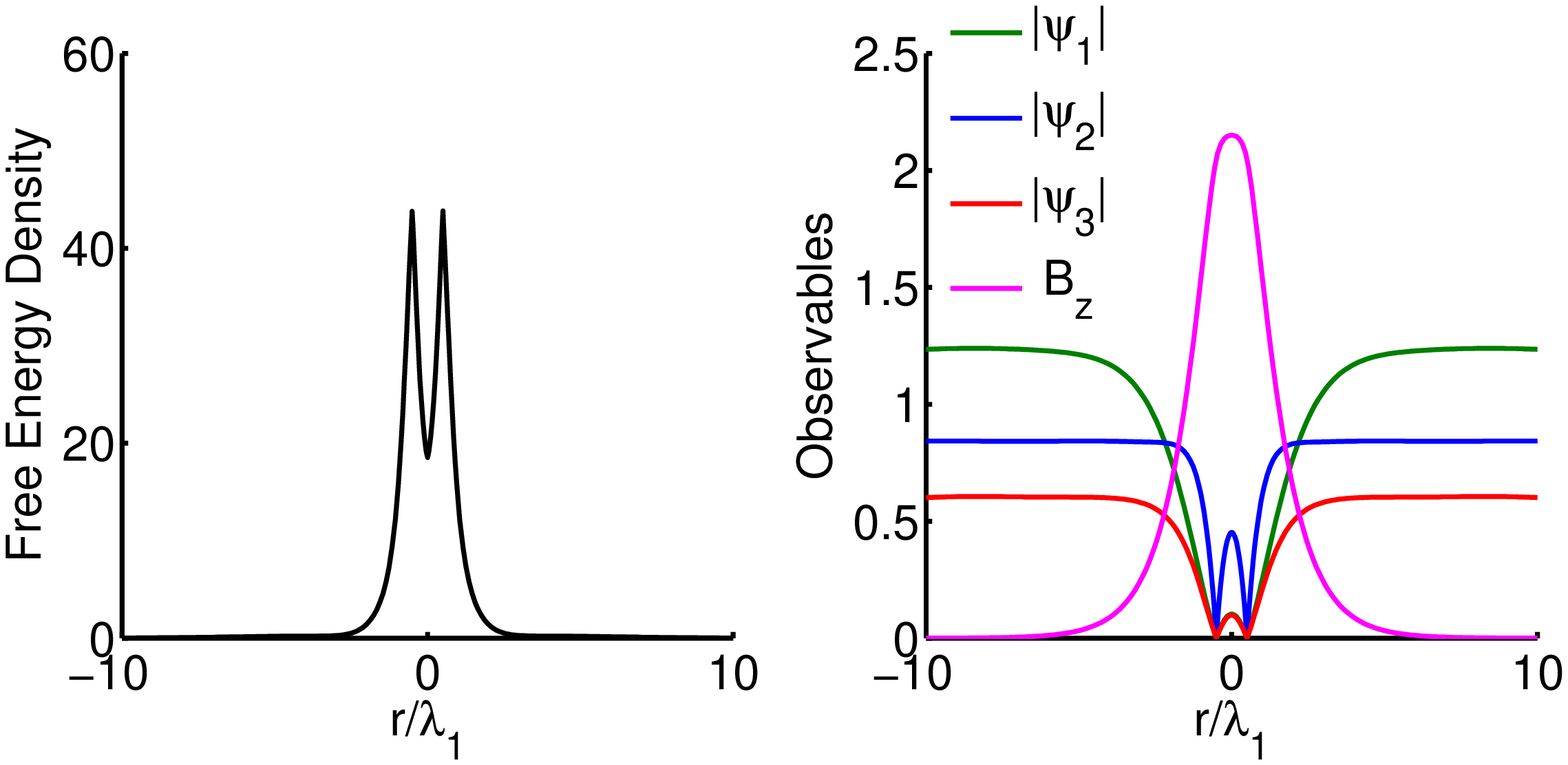} 
\label{threeOtwoco1} 
} 
\\ 
\subfigure{$d=2$}\\ 
{
\includegraphics[width=1\linewidth]{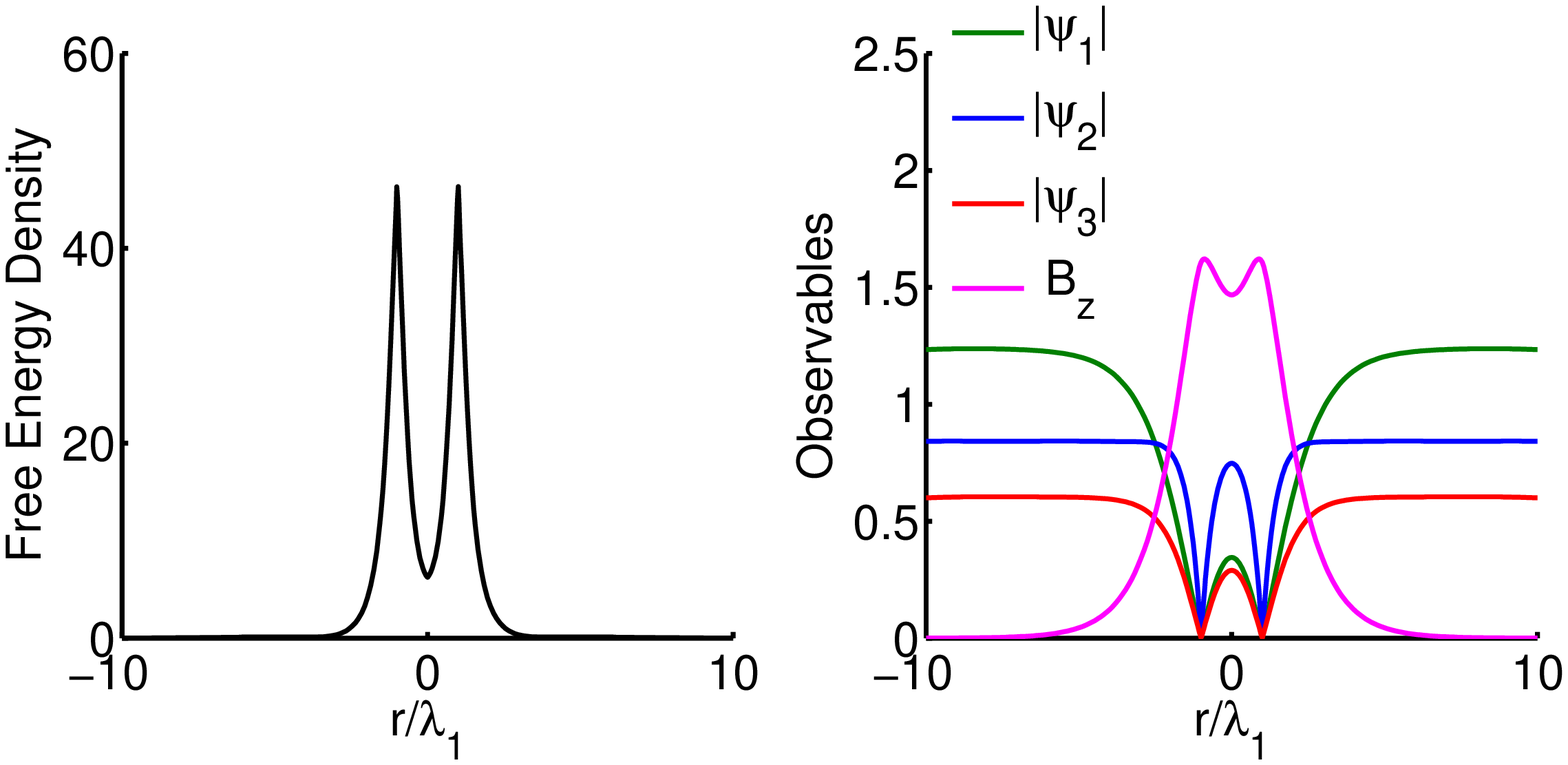} 
\label{threeOtwoco2} 
} 
\\ 
\subfigure{$d=3$}\\ 
{
\includegraphics[width=1\linewidth]{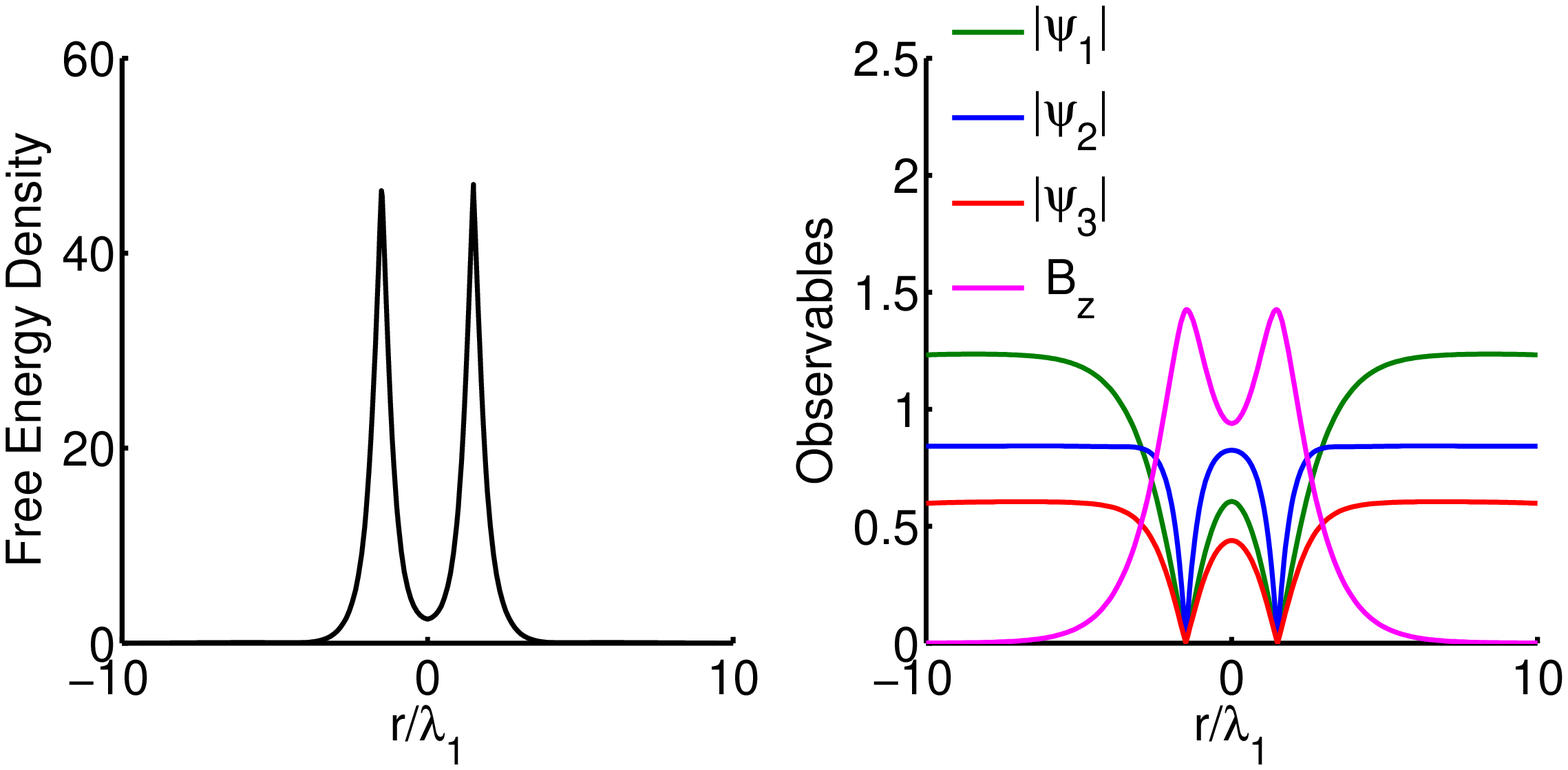} 
\label{threeOtwoco3} 
} 
\\ 
\subfigure{$d=4$}\\ 
{
\includegraphics[width=1\linewidth]{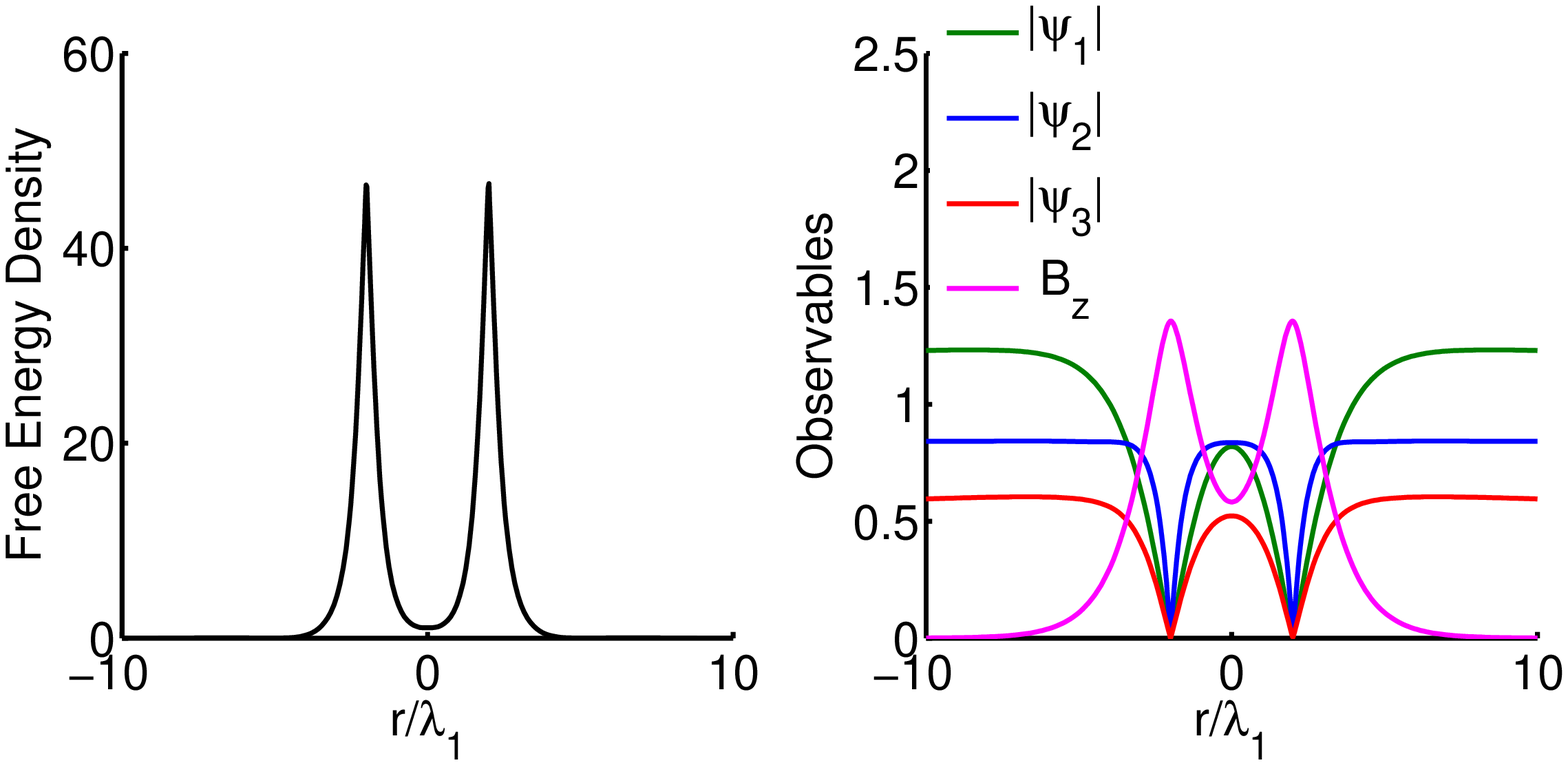} 
\label{threeOtwoco4} 
} 
\end{minipage} 
\begin{minipage}[b]{0.5\linewidth} 
\centering 
\subfigure{$d=5$}\\ 
{ 
\includegraphics[width=1\linewidth]{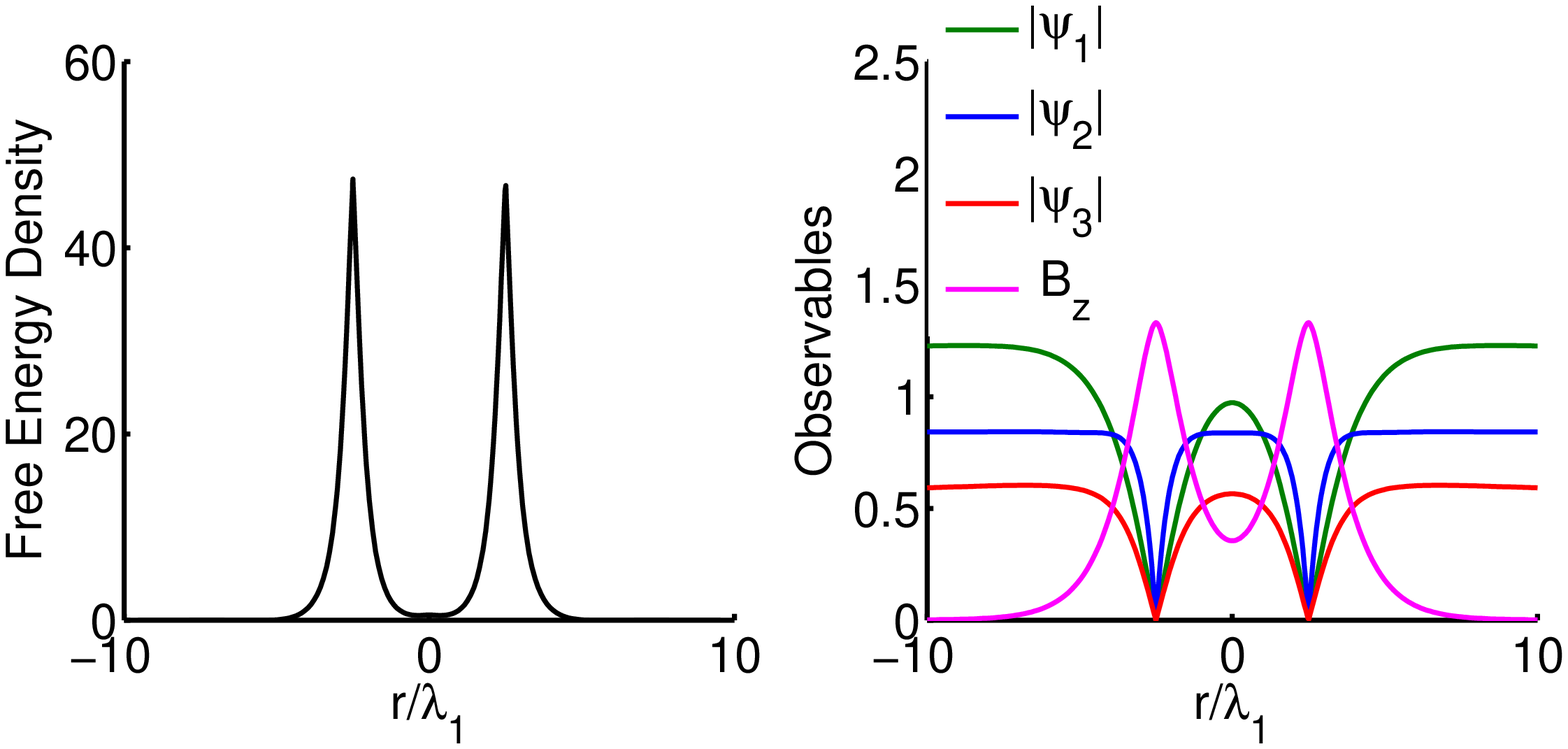} 
\label{threeOtwoco5} 
} 
\\ 
\subfigure{$d=6$}\\ 
{ 
\includegraphics[width=1\linewidth]{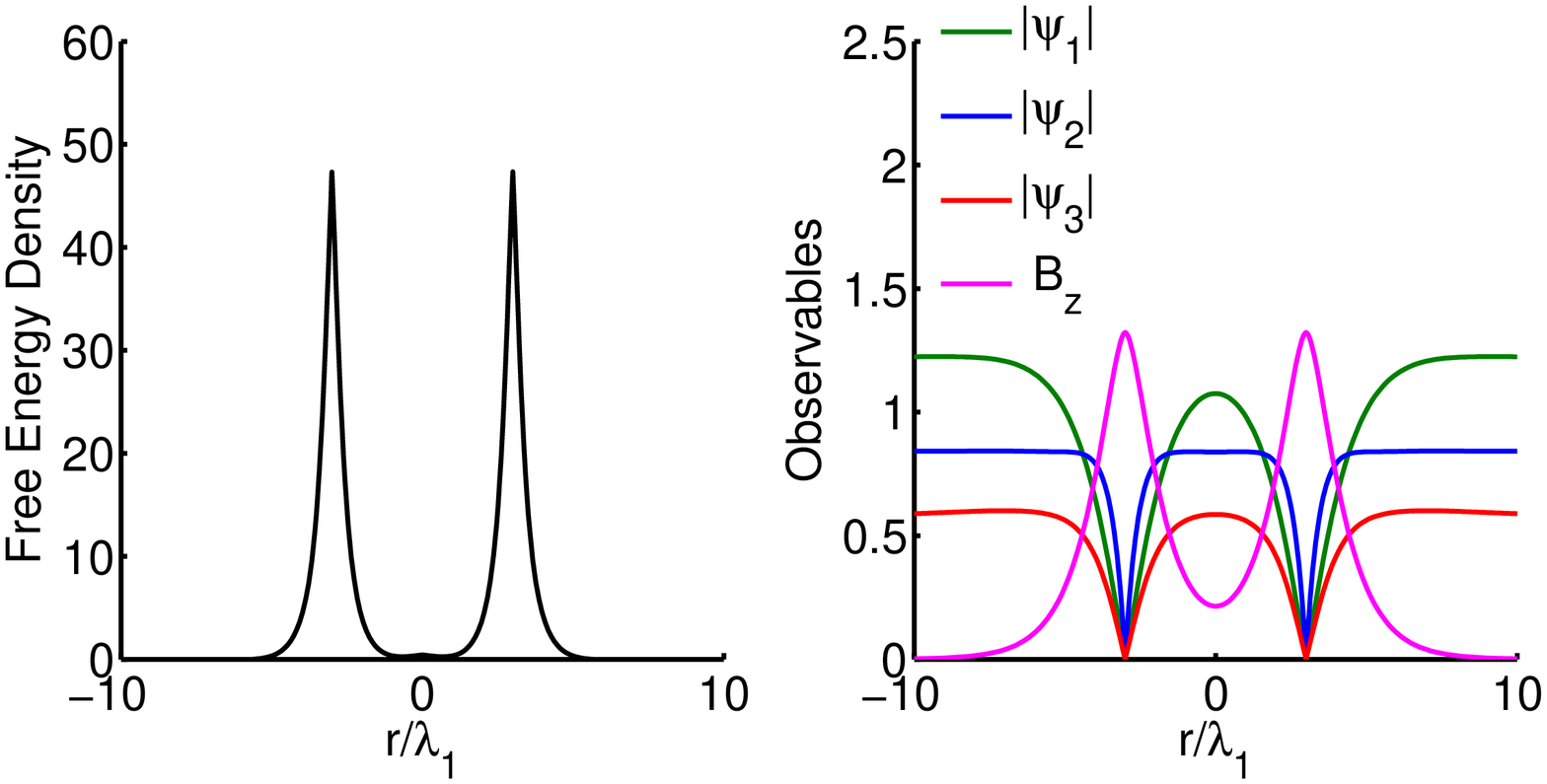} 
\label{threeOtwoco6} 
} 
\\ 
\subfigure{$d=7$}\\ 
{
\includegraphics[width=1\linewidth]{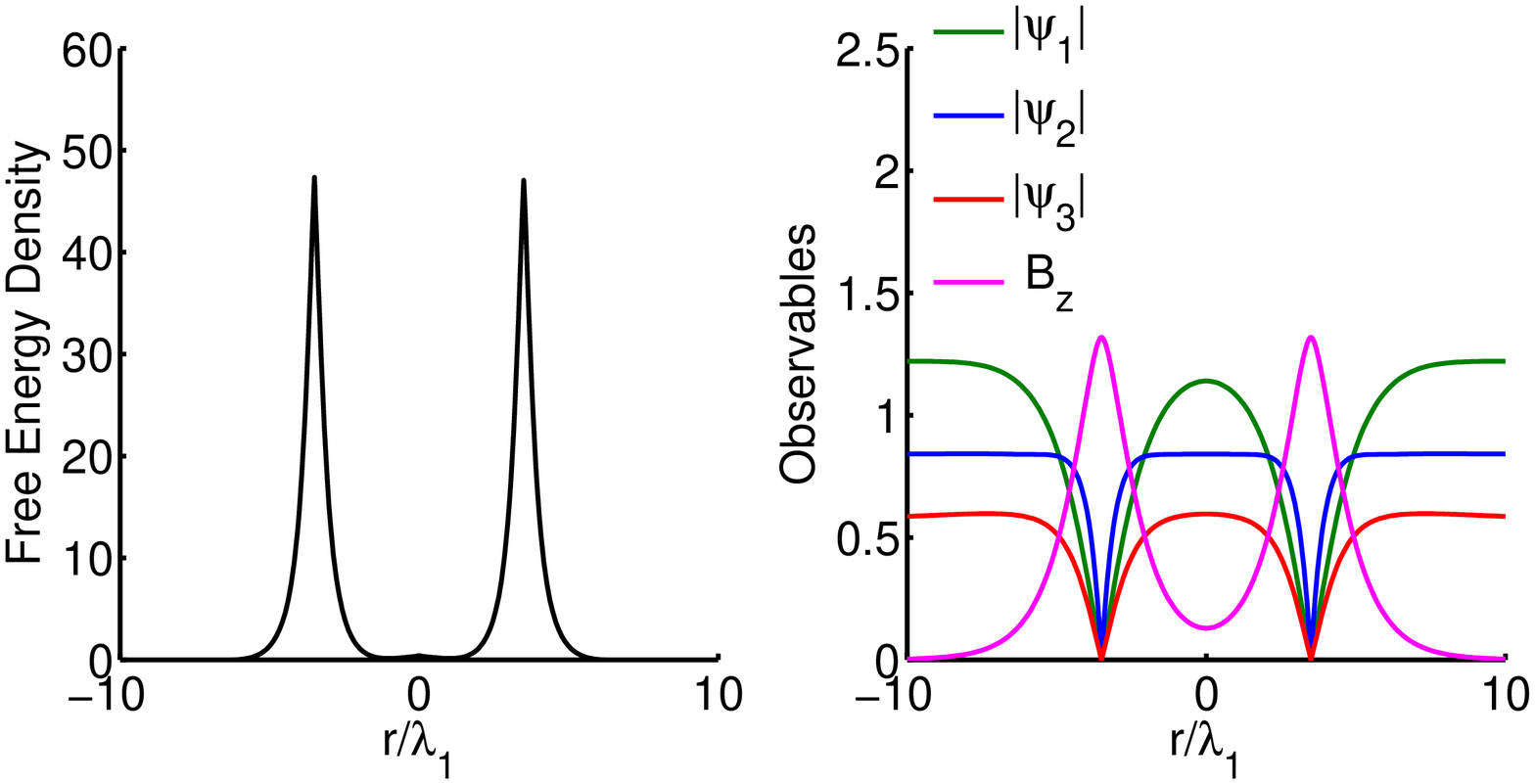} 
\label{threeOtwoco7} 
} 
\\ 
\subfigure{($ d=8 $}\\ 
{
\includegraphics[width=1\linewidth]{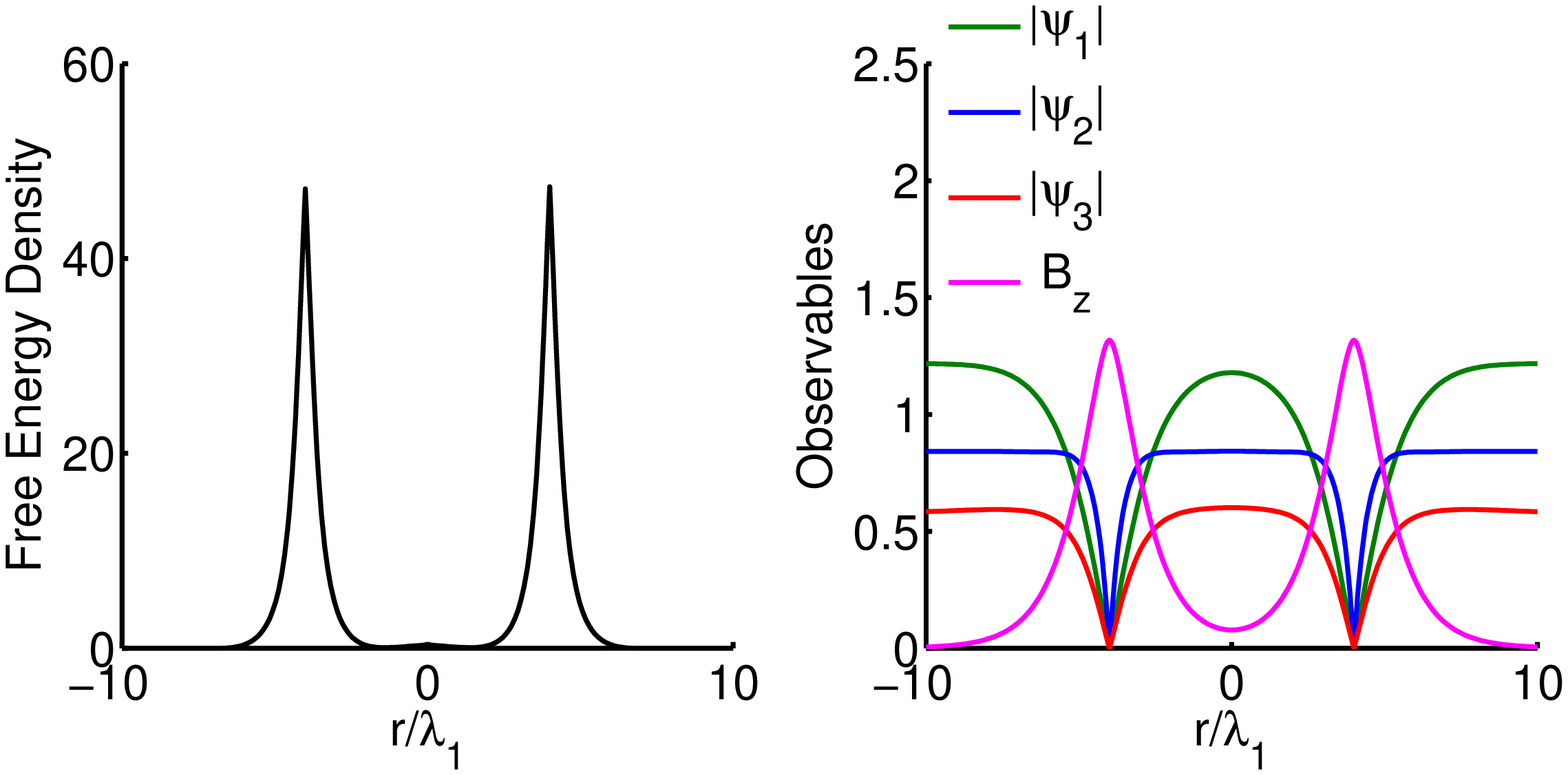} 
\label{threeOtwoco8} 
} 
\end{minipage} 

\caption{Free energy density and condensation states and magnetic field profiles cross sections in a plane for two vortices with three condensations at different 
distances using the variational method. As the distance increases magnetic field profile function decreases  between vortices. When vortices 
are close to each other the magnetic field profile function shows an increase. Repulsion between vortices happens as the result of the increase of magnetic 
field. A change of the behavior of the first condensation relative to the other condensations from the distance $d=1$ up to the distance $d=4$ is observed. This shows that the rate of change of energy versus the distance between two vortices at such distances are not monotonic.} 
\label{En3}
\end{figure} 
 
\begin{figure}
\centering 
\includegraphics[width=0.5\linewidth]{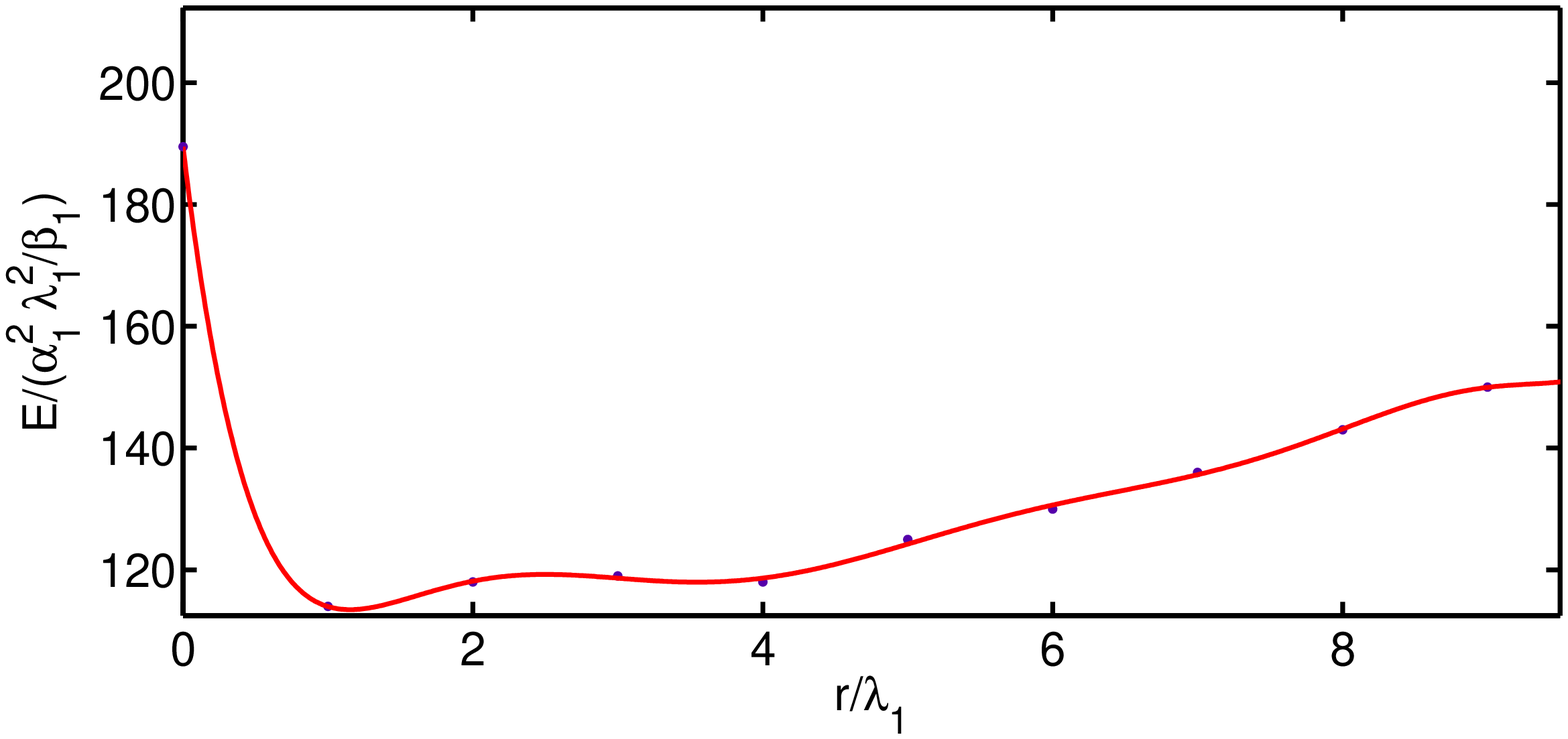} 
\caption{Interaction energy dependence on distance between two vortices with three condensation states obtained by the variational method. 
There are two stability points at distances $1$ and $4$. At large distances there is attraction between these vortices. At small distances the 
interaction is repulsion. Two stability points lead to more complex structure for the location of vortices in these materials. Also it may have some 
novel practical usage} 
\label{ED3}
\end{figure} 
We plot the energy versus distance by the same method as the previous section. Figure \ref{En3} shows the free energy density and vortex profiles of two vortices at different distances. Figure \ref{ED3} shows the  free energy of two vortices versus distances. Again there 
is a repulsion between two vortices at short distances and attraction when they are far from each other. There are two stability points for these 
vortices, one in $4$ and the other at $1$. So, three condensations states can be different with respect to the two condensation states: the number 
of stability points is increased and the energy of the vortex formation increases compared with the superconductors of type I and II and $1.5$. Note that the presence of two stable points depends on the value of the parameters of the model. One could consider the correlation lengths and penetration depths relative to each other such that it leads to only one stable point. 
  
Because of the existence of two stability points, quantum tunneling may occur for the system of two vortices  between the two stability points if one considers the time in the calculations. This may lead to the possibility of the existence of another topological structure, such as an instanton, in the superconductor materials with three condensation states. The possibility of the existence of Skyrme structures in these materials has recently been studied theoretically \cite{Garaud2014}. This evidence suggests that the Ginzburg-Landau Lagrangian with three condensations has theoretical properties which do not have an analog in the ordinary superconductors. 

\section{conclusion} 
We use a numerical method to obtain the vortex profiles and the interaction between the vortices for a three condensation state superconductor. 
In this method, we use some trial functions for condensations and  the magnetic field. The variational parameters of these functions are obtained by 
minimizing the free energy. We calculate  the free energy density integral which is the energy of vortex formation in a polar coordinate system. The 
energy of a vortex with three condensations is higher than the two condensation states. The energy of two condensations is also larger than for 
type I and II superconductors with the same penetration depth and correlation length. 
Since these materials with two and three condensations are high temperature superconductors, it might be a hint that there is a relation between
the energy of this structure and the higher phase transition temperature \cite{Klienert} in this type of superconductor. 
We have figured out that there are different types  of interactions between these vortices: In type I and II superconductors, the interaction energy 
of a vortex with winding $n=2$ and two vortices with $n=1$ can show the type of interaction when they are far from each other. We have obtained 
attraction for type I and repulsion for type II superconductors. Using a full procedure of the variational method for a type $1.5$ superconductor in a polar 
coordinate system, we obtain repulsion at smaller distances than $2.7/\lambda_{1}$ and attraction at larger distances. There is a stability point for vortices at $2.7\lambda_{1}$ 
in this case.
For three condensations, we have seen the same behavior as the two condensations; but there are two stability points at $4$ and one. 

Currents and magnetic fields lead to a repulsion type of interaction and also the core of the condensation can lead to an attraction type of interaction when $r \gg 1$  \cite{Speight2005}. 
For type I where $\frac{\lambda}{\xi}<\frac{1}{\sqrt{2}}$, the core of the magnetic field is smaller than the core of the condensation. Thus, the winner of the interaction is attraction \cite{Kramer,Peeters2011}. For type II the situation is reversed and a repulsion interaction exists. 
A type 1.5 superconductor with $\xi_{1} \ll \lambda_{1}$ and $\lambda_{2}\ll \xi_{2}$ can be considered as superconductor of type II according to the $\xi_{1},\lambda_{1}$  and a superconductor of type I according to $\xi_{2},\lambda_{2}$. The size of the core of one of the components is the largest length scale of the problem. Therefore a region domination of the repulsive interaction mediated by currents and magnetic field and a region of domination of the attraction mediated by the largest length scale of the problem exist. A schematic view of this type of superconductor is illustrated in \cite{Speight2005}. The stability point is at the border of these two regions.
A superconductor with three condensations with $\xi_{1} \gg \lambda_{1}$ and $\lambda_{2}\gg \xi_{2}$ and $\lambda_{3}\gg \xi_{3}$ can be considered as two type 1.5 superconductors. $\xi_{1} \gg \lambda_{1}$ and $\lambda_{2}\gg \xi_{2}$ represent a superconductor of type 1.5 with a stability point at $2.7 \lambda_{1}$. $\xi_{1} \gg \lambda_{1}$ and $\lambda_{3}\gg \xi_{3}$ represents  another type 1.5 with a stability point at another location. When all of these length scales are present there is competition between these two type 1.5 superconductors. This may lead to the existence of two stability points. There exists an effective penetration depth for large distances.
This length is obtained from the London approximation. The effective penetration length will be important when the
gradients of the condensations are negligible. This happens when $r$ is
in the region where all the condensations obtain their asymptotic values. However, the situation is different for smaller distances. Three individual penetration depths are  introduced 
because of the response of the magnetic field to each condensation. 
The competition between repulsion given by penetration depths and attractive mechanisms given 
by condensations changes the monotonic behavior of the energy for three condensation superconductors, especially for intermediate distances (Fig. \ref{ED3}).  However, the interband coupling and the nonlinearity of the equations make the system more complex than the above simple description. So the number of stable points depends on the values of these three correlation lengths and penetration depths of the model. Here we use a penetration depth which conveys all other lengths of the model. One could use parameters that do not lead to such a system with two stable point.    
The existence of two stable points may have novel applications. Because of the existence of two stability points, quantum tunneling may occur for the system of two vortices  between the two stability points if one considers the time in the calculations. This may lead to the possibility of the existence of another topological structure, such as  the instanton in the superconductor materials with three condensation states. The possibility of the existence of Skyrme structures in these materials has recently been studied theoretically \cite{Garaud2014}. Recent experimental observation on the vortex behavior in these type of materials, which can be described with three condensations, have been shown to have different behavior of the vortices \cite{Philip Moll}. This evidence suggests that the Ginzburg-Landau Lagrangian with three condensations has theoretical properties which do not have an analog in ordinary superconductors. 
If the energy of these structures has something to do with the temperature, then theoretically we can predict what values of the correlation lengths
and penetration depths lead to a higher energy for the vortex formation and therefore a higher phase transition temperature. If, seen from 
the experimental point of view, making or finding such materials with these penetration depths and correlation lengths is made possible,
 higher critical temperature than the current ones can be reachable. It may be possible to apply this numerical method to study the interaction between special types of non- abelian vortices. 
 
\section{Acknowledgement} 
We are grateful to the research council of the University of Tehran for supporting this study.


\end{document}